\documentclass[11pt,a4paper]{article}
\pdfoutput=1
\usepackage{jheppub}
\usepackage{mathtools,bm,placeins,microtype}
\usepackage[nameinlink,capitalise]{cleveref}
\numberwithin{equation}{section}
\crefname{equation}{Eqn.}{Eqns.}
\Crefname{equation}{Eqn.}{Eqns.}
\crefname{section}{Sec.}{Secs.}
\Crefname{section}{Sec.}{Secs.}
\crefname{subsection}{Sec.}{Secs.}
\Crefname{subsection}{Sec.}{Secs.}
\crefname{appendix}{App.}{Apps.}
\Crefname{appendix}{App.}{Apps.}
\crefname{subappendix}{App.}{Apps.}
\Crefname{subappendix}{App.}{Apps.}
\crefname{figure}{Fig.}{Figs.}
\Crefname{figure}{Fig.}{Figs.}

\newcommand{\be}{\begin{equation}}
\newcommand{\ee}{\end{equation}}
\newcommand{\ii}{\mathrm{i}}
\newcommand{\dd}{\mathrm{d}}
\newcommand{\Tr}{\operatorname{Tr}}
\newcommand{\Log}{\operatorname{Log}}
\newcommand{\sgn}{\operatorname{sgn}}
\newcommand{\one}{\mathbf1}
\newcommand{\ket}[1]{\lvert#1\rangle}
\newcommand{\bra}[1]{\langle#1\rvert}
\newcommand{\braket}[2]{\langle#1\vert#2\rangle}
\newcommand{\proj}[1]{\lvert#1\rangle\langle#1\rvert}
\newcommand{\eps}{\varepsilon}
\newcommand{\cD}{\mathcal D}
\newcommand{\SR}{S^{\mathrm R}}
\newcommand{\SI}{S^{\mathrm I}}
\newcommand{\PaperTitle}{Pseudo entropy from entanglement entropy}
\newcommand{\PaperPreprint}{YITP-26-124}
\newcommand{\PaperAbstract}{Pseudo entropy extends entanglement entropy from a single quantum state to a pair of nonorthogonal states and is generally complex. Taking advantage of Cauchy-Riemann equations, Kramers-Kronig relations, and analytic continuation in state parameters, we show how and to what extent real and imaginary parts of pseudo entropy can be derived from ordinary entanglement entropy. For families with holomorphic coefficients in finite-dimensional Hilbert spaces, the reduced transition matrix equals the ordinary reduced density matrix formula evaluated at complex parameters. Then a convergent Taylor series gives the real and imaginary parts of pseudo entropy from even and odd derivatives of entanglement entropy at the real midpoint. The matrix identity also gives formulae for excess pseudo entropy and R\'enyi entropies, and interpolation formulae for families with polynomial coefficients. We apply these results to boundary-state quenches and thermal states in conformal field theory, and to fermionic and bosonic Gaussian states and quenches. In conformal field theory, Kramers-Kronig relations give the first moment of the imaginary part in terms of the central charge and one-point functions for boundary states.}
\preprint{\PaperPreprint}
\title{\PaperTitle}
\hypersetup{ pdftitle={\PaperTitle}, pdfauthor={Abhigyan Saha, Piotr Sułkowski, Tadashi Takayanagi}, pdfkeywords={Pseudo entropy, Entanglement entropy, Analytic continuation, Quantum quenches} }
\author[a]{Abhigyan Saha,}
\author[a]{Piotr Su\l{}kowski,}
\author[b,c]{Tadashi Takayanagi}
\affiliation[a]{Faculty of Physics, University of Warsaw, Pasteura 5, 02-093 Warsaw, Poland}
\affiliation[b]{Center for Gravitational Physics and Quantum Information, Yukawa Institute for Theoretical Physics, Kyoto University, Kitashirakawa Oiwakecho, Sakyo-ku, Kyoto 606-8502, Japan}
\affiliation[c]{Inamori Research Institute for Science, 620 Suiginya-cho, Shimogyo-ku, Kyoto 600-8411, Japan}
\abstract{\PaperAbstract}
\begin{document}
\maketitle
\flushbottom

\section{Introduction}
For bipartite pure states, entanglement entropy quantifies entanglement in quantum many-body systems and quantum field theory \cite{Horodecki:2009Entanglement,CalabreseCardy:2004}. For static states in holographic conformal field theories (CFTs), the holographic entanglement prescription relates it to the area of a bulk minimal surface \cite{RyuTakayanagi:2006}, a relation central to work on the emergence of spacetime from quantum information \cite{Takayanagi:2025Essay}.

For two nonorthogonal states, pseudo entropy is defined from their reduced transition matrix \cite{Nakata:2020luh} for a subsystem. It equals entanglement entropy when the states coincide and is generally complex when they differ. Timelike entanglement entropy was introduced in CFTs and related by analytically continuing the originally spacelike subsystem into a timelike one \cite{Doi:2022TimelikePRL,Doi:2023Timelike}. Via the AdS/CFT \cite{Maldacena:1997re}, we can find geometric interpretations of the timelike entanglement entropy as the complex valued areas of extremal surfaces \cite{Doi:2022TimelikePRL,Doi:2023Timelike,Heller:2024whi,Heller:2025kvp}, motivating proposals relating its imaginary part to the emergence of time \cite{Takayanagi:2025Essay}. Related work connects timelike entanglement to causal influences under unitary evolution and to traversable AdS wormholes \cite{Harper:2025timelike}, to spacelike entanglement entropy and its first-order temporal derivative \cite{GuoHeZhang:2024TimelikeSpacelike}, and to higher-order temporal derivatives of twist operators \cite{XuGuo:2025Imaginary}. See also e.g. \cite{Narayan:2022afv,Foligno:2023dih,Narayan:2023ebn,Chu:2023zah,Grieninger:2023knz,Jena:2024tly,Milekhin:2025ycm,Chu:2025sjv,Bou-Comas:2024pxf,Nunez:2025gxq,Nunez:2025puk,Fujiki:2025rtx} for other related work.

The entropy of normalised singular values \cite{Roga:2013receiver} gives SVD entanglement entropy for a reduced transition matrix \cite{Parzygnat:2023SVD}. In Chern-Simons theory, the imaginary part of pseudo entropy can detect link chirality \cite{Caputa:2024SVD}, while a pseudo-metric interpretation of the absolute value of the excess SVD entropy was obtained for two-component \(U(1)\) link states \cite{Caputa:2024SVD}. The entanglement entropies based on Unit-Invariant Singular Value Decomposition (UISVD) were introduced in \cite{Caputa:2025UISVD}. Earlier formulae relate reduced transition matrices to reduced density matrices of superposition states and give relations for pseudo-R\'enyi entropies \cite{GuoZhang:2023sumrule}. Related formulae follow by analytically continuing the coefficients of these states \cite{GuoJiangXu:2024}. The Kramers-Kronig relations between the real and imaginary parts of thermal pseudo entropy \cite{Caputa:2024thermal} motivate the question below.

We ask under what conditions the real and imaginary parts of pseudo entropy can be derived from the parameter dependence of ordinary entanglement entropy. In a finite-dimensional Hilbert space, for two nonorthogonal states in the same family with holomorphic coefficients, the reduced transition matrix equals the ordinary reduced density matrix formula at a complex parameter value fixed by the midpoint and half-difference of the two parameter values. If the normalisation remains nonzero and the spectrum of the reduced matrix stays in the domain of the chosen logarithm branch on a complex neighbourhood of the parameter values used in the continuation, the entropy is holomorphic there. It then has derivatives of every order and a local Taylor series. When this series converges at the required complex value, pseudo entropy is given by derivatives of ordinary entanglement entropy evaluated at the real midpoint. Even and odd derivatives give its real and imaginary parts. The entanglement entropies of the two chosen states alone are insufficient.

\Cref{sec-entropy-continuation} gives the Cauchy-Riemann and Taylor relations. \Cref{sec-finite-families} contains the matrix identity and the resulting entropy formulae. The section also includes excess pseudo entropy, R\'enyi entropies and interpolation from finitely many values for families with polynomial coefficients. \Cref{sec-cft} applies the continuation to boundary-state quenches and thermal states, giving integral relations for the imaginary part. \Cref{sec-gaussian} gives the corresponding formulae for fermionic and bosonic Gaussian states and identifies when quench transition matrices can be written at complex times. The appendices contain the proofs and the conditions needed for the continuations and limits.

\section{Entanglement entropy and analytic continuation}\label{sec-entropy-continuation}
All matrices in this section act on finite-dimensional Hilbert spaces. For positive real numbers, \(\log\) is the real logarithm, while \(\Log\) is the principal logarithm on \(\mathbb C\setminus(-\infty,0]\). For a matrix \(M\) with spectrum in this set, \(\Log M\) denotes its principal matrix logarithm. The matrix need not be diagonalisable, as shown in \cref{app-holomorphy-and-taylor-series}. All normalisation denominators are nonzero, and every matrix to which \(\Log\) is applied is invertible.

Let \(\mathcal H=\mathcal H_A\otimes\mathcal H_B\). For a nonzero pure state \(\ket{\psi}\), define
\be
\rho\coloneqq\frac{\ket{\psi}\bra{\psi}}{\braket{\psi}{\psi}},\qquad\rho_A\coloneqq\Tr_B\rho,\qquad S^{\mathrm E}(\psi)\coloneqq-\Tr[\rho_A\log\rho_A]=-\sum_jp_j\log p_j,
\ee
where \(p_j\) are the eigenvalues of \(\rho_A\), and we use the convention \(0\log0=0\). For \(\braket{\psi_2}{\psi_1}\neq0\), define \cite{Nakata:2020luh}
\be
\tau\coloneqq\frac{\ket{\psi_1}\bra{\psi_2}}{\braket{\psi_2}{\psi_1}},\qquad\tau_A\coloneqq\Tr_B\tau,\qquad S(\tau_A)\coloneqq-\Tr\!\left[\tau_A\Log\tau_A\right].
\ee
The transition matrix has trace one and is equal to \(\rho\) when the states coincide. For orthogonal states the overlap vanishes and this normalised transition matrix is undefined. We fix subsystem \(A\) throughout and trace out \(B\). The same formulae apply to \(B\), with \(\Tr_B\) replaced by \(\Tr_A\).

Let \(\rho_A(\bm z)\) be a matrix-valued function holomorphic on \(\mathcal U\subset\mathbb C^d\) whose value at \(\bm z=\bm x\in\mathbb R^d\) is the reduced density matrix \(\rho_A(\bm x)\). At complex parameter values it need not be Hermitian or positive. The values of \(\rho_A(\bm x)\) on a nonempty open subset of \(\mathbb R^d\) uniquely fix the holomorphic matrix-valued function on the connected domain \(\mathcal U\), as shown in \cref{app-holomorphy-and-taylor-series}. Write \(\SR\coloneqq\operatorname{Re}S\) and \(\SI\coloneqq\operatorname{Im}S\). With \(\bm z=\bm x+\ii\bm y\), define
\begin{align}
&S(\bm z) \coloneqq-\Tr[\rho_A(\bm z)\Log\rho_A(\bm z)] = \SR(\bm x,\bm y)+\ii\SI(\bm x,\bm y), \\ &S^{\mathrm E}(\bm x) \coloneqq-\Tr[\rho_A(\bm x)\log\rho_A(\bm x)] = S(\bm x) = \SR(\bm x,\bm0), \qquad\SI(\bm x,\bm0) =0.
\end{align}
By \cref{app-holomorphy-and-taylor-series}, \(S\) is holomorphic under these assumptions. For each component \(z^a\), \(a=1,\ldots,d\), fix \(z^b\) for every \(b\neq a\) and apply the one-variable Cauchy-Riemann equations to \(z^a\mapsto S(z^1,\ldots,z^d)\). Writing \(\partial_{x^a}=\partial/\partial x^a\) and \(\partial_{y^a}=\partial/\partial y^a\), one obtains
\be
\partial_{y^a}\SR=-\partial_{x^a}\SI,\qquad\partial_{y^a}\SI=\partial_{x^a}\SR.\label{eq-cauchy-riemann-equations}
\ee
A series of derivatives at real parameters gives the entropy at a complex parameter value only when the series converges there. Set \(g(s)\coloneqq S(\bm x+\ii s\bm y)\) and define \(\cD_{\bm y}\coloneqq\sum_{a=1}^dy^a\partial_{x^a}\). Assume that \(\bm x+\ii s\bm y\in\mathcal U\) for every real \(0\leq s\leq1\), with the matrix and logarithm assumptions above valid on a neighbourhood of this segment. If the Taylor series of \(g\) about \(s=0\) converges at \(s=1\), then
\be
S(\bm x+\ii\bm y)=\sum_{n=0}^{\infty}\frac{\ii^n}{n!}\cD_{\bm y}^{\,n}S^{\mathrm E}(\bm x)=e^{\,\ii\cD_{\bm y}}S^{\mathrm E}(\bm x).\label{eq-entropy-taylor-series}
\ee
A sufficient convergence condition is that \(\bm x+\ii s\bm y\in\mathcal U\) for \(|s|<r_s\), with \(r_s>1\), and that the matrix and logarithm assumptions above hold throughout this disc. Holomorphy and the chain rule give
\be
g^{(n)}(0)=\ii^n\left(\sum_{a=1}^dy^a\partial_{z^a}\right)^nS(\bm x)=\ii^n\left(\sum_{a=1}^dy^a\partial_{x^a}\right)^nS^{\mathrm E}(\bm x)=\ii^n\cD_{\bm y}^{\,n}S^{\mathrm E}(\bm x).
\ee
Taylor's theorem \cite{Orloff:2018topic7} gives \cref{eq-entropy-taylor-series} under this sufficient condition. The proof under the more general assumptions on the segment and convergence is given in \cref{app-holomorphy-and-taylor-series}. Sufficient local conditions for convergence are given below for finite-dimensional and bosonic Gaussian states. Separating the even and odd powers gives
\be
\SR(\bm x,\bm y)=\cos(\cD_{\bm y})S^{\mathrm E}(\bm x),\quad\SI(\bm x,\bm y)=\sin(\cD_{\bm y})S^{\mathrm E}(\bm x),\quad\SI(\bm x,\bm y)=\tan(\cD_{\bm y})\SR(\bm x,\bm y).\label{eq-real-imaginary-taylor-series}
\ee
The last equality uses \(\tan(\cD_{\bm y})\coloneqq\sin(\cD_{\bm y})[\cos(\cD_{\bm y})]^{-1}\) on the range of \(\cos(\cD_{\bm y})\), for a common class of functions on which both the sine and cosine series converge and the cosine is one-to-one. No Taylor series for the tangent is assumed. For one complex variable, \cref{app-integral-formulae} gives an integral formula on a horizontal strip.

We use \cref{eq-real-imaginary-taylor-series} below for the real and imaginary parts. \Cref{sec-finite-families} identifies the complex parameter value at which the matrix equals the reduced transition matrix of two states.

\section{Transition matrices from states with holomorphic coefficients}\label{sec-finite-families}
For states in finite-dimensional Hilbert spaces whose coefficients are holomorphic functions of common parameters, the formula below is an ordinary density matrix at real parameter values and the transition matrix of two chosen states at a complex parameter value. Under the assumptions stated below for each quantity, this gives formulae for pseudo entropy and pseudo-R\'enyi entropy. We then compare the real part with the average of the two entanglement entropies, derive the excess, and give interpolation formulae for state families with polynomial coefficients. The assumptions of \cref{sec-entropy-continuation} apply throughout.

\subsection{Two states with holomorphic coefficients}\label{sec-two-states}
In a fixed orthonormal basis of \(\mathcal H=\mathcal H_A\otimes\mathcal H_B\), let
\be
\ket{\psi(\bm z)}\coloneqq\sum_nc_n(\bm z)\ket{n},\qquad\bm z\in\mathbb C^d,\qquad\widetilde c_n(\bm z)\coloneqq\overline{c_n(\overline{\bm z})}.
\ee
The sum is finite, the vector is nonzero, and both \(c_n\) and \(\widetilde c_n\) are holomorphic. For real \(\bm z\), \(\widetilde c_n(\bm z)=\overline{c_n(\bm z)}\). The combination \(\overline{c_n(\overline{\bm z})}\) is also used in \cite{Harper:2025timelike}. For real \(\bm t,\bm\alpha\), write \(\bm z=\bm\alpha+\ii\bm t\). The coefficients in the ket and bra then have arguments \(\bm\alpha+\ii\bm t\) and \(\bm\alpha-\ii\bm t\), respectively. We use complex \(\bm t\) and \(\bm\alpha\) so that these two arguments can be chosen independently, giving different states in the ket and bra. For \((\bm t,\bm\alpha)\in\mathbb C^d\times\mathbb C^d\), define
\be
\rho(\bm t,\bm\alpha)\coloneqq\frac{\left[\sum_nc_n(\bm\alpha+\ii\bm t)\ket{n}\right]\left[\sum_m\widetilde c_m(\bm\alpha-\ii\bm t)\bra{m}\right]}{\sum_\ell\widetilde c_\ell(\bm\alpha-\ii\bm t)c_\ell(\bm\alpha+\ii\bm t)}.\label{eq-holomorphic-matrix-family}
\ee
The denominator is the trace of the numerator, so \(\Tr\rho=1\) when it is nonzero, and every matrix entry is holomorphic in \((\bm t,\bm\alpha)\). For real \((\bm t,\bm\alpha)\), this becomes the ordinary density matrix of \(\ket{\psi(\bm\alpha+\ii\bm t)}\). Set
\be
\rho_A(\bm t,\bm\alpha)\coloneqq\Tr_B\rho(\bm t,\bm\alpha),\quad S^{\mathrm E}(\bm t,\bm\alpha)\coloneqq-\Tr\!\left[\rho_A(\bm t,\bm\alpha)\log\rho_A(\bm t,\bm\alpha)\right],\quad(\bm t,\bm\alpha)\in\mathbb R^d\times\mathbb R^d.
\ee
Choose two real parameter pairs \((\bm t_i,\bm\alpha_i)\), \(i=1,2\), \(\ket{\psi_i}\coloneqq\ket{\psi(\bm\alpha_i+\ii\bm t_i)}\), and define
\be
\bm t_{\mathrm c}\coloneqq\frac{\bm t_1+\bm t_2}{2},\qquad\bm t_\delta\coloneqq\frac{\bm t_2-\bm t_1}{2},\qquad\bm\alpha_{\mathrm c}\coloneqq\frac{\bm\alpha_1+\bm\alpha_2}{2},\qquad\bm\alpha_\delta\coloneqq\frac{\bm\alpha_2-\bm\alpha_1}{2}.\label{eq-parameter-centre-and-half-difference}
\ee
Substituting \((\bm t,\bm\alpha)=(\bm t_{\mathrm c}+\ii\bm\alpha_\delta,\bm\alpha_{\mathrm c}-\ii\bm t_\delta)\) in \cref{eq-holomorphic-matrix-family} gives \(\ket{\psi_1}\bra{\psi_2}\) in the numerator and \(\braket{\psi_2}{\psi_1}\) in the denominator. Taking the partial trace therefore gives the reduced transition matrix.
\be
\bm\alpha+\ii\bm t=\bm\alpha_1+\ii\bm t_1,\qquad\bm\alpha-\ii\bm t=\bm\alpha_2-\ii\bm t_2,\qquad\tau_A=\rho_A(\bm t_{\mathrm c}+\ii\bm\alpha_\delta,\,\bm\alpha_{\mathrm c}-\ii\bm t_\delta).\label{eq-transition-matrix-complex-parameters}
\ee
\cite{GuoJiangXu:2024} uses a related analytic continuation. Here \cref{eq-transition-matrix-complex-parameters} identifies the reduced transition matrix with the analytic continuation of the density-matrix formula for every family of states in a finite-dimensional Hilbert space satisfying the assumptions above.

Relative to \((\bm t_{\mathrm c},\bm\alpha_{\mathrm c})\), the complex parameter change in \cref{eq-transition-matrix-complex-parameters} is \(\ii(\bm\alpha_\delta,-\bm t_\delta)\). The derivatives in the Taylor series are evaluated at the real midpoint, where \(\rho_A\) is an ordinary reduced density matrix and its entropy is \(S^{\mathrm E}\). Define \(\cD_\perp\) on the real variables \((\bm t,\bm\alpha)\) by
\be
\cD_\perp\coloneqq\sum_{a=1}^d\left(\alpha_\delta^a\partial_{t^a}-t_\delta^a\partial_{\alpha^a}\right).\label{eq-D-perp-definition}
\ee
Under the assumptions for the Taylor expansion in \cref{eq-entropy-taylor-series}, applied separately to the reduced matrix and its entropy, one obtains
\be
\tau_A=\left.e^{\ii\cD_\perp}\rho_A(\bm t,\bm\alpha)\right|_{\bm t=\bm t_{\mathrm c},\,\bm\alpha=\bm\alpha_{\mathrm c}},\qquad S(\tau_A)=\left.e^{\ii\cD_\perp}S^{\mathrm E}(\bm t,\bm\alpha)\right|_{\bm t=\bm t_{\mathrm c},\,\bm\alpha=\bm\alpha_{\mathrm c}}.\label{eq-transition-matrix-and-entropy-taylor-series}
\ee
The identity \(\rho_A(\bm t,\bm\alpha)^\dagger=\rho_A(\overline{\bm t},\overline{\bm\alpha})\) implies that the even Taylor coefficients are Hermitian and the odd coefficients are anti-Hermitian, as shown in \cref{app-holomorphy-and-taylor-series}. More generally, suppose that \(M\mapsto\mathcal F(M)\) is scalar- or matrix-valued and holomorphic on an open set containing \(\rho_A(\bm t_{\mathrm c}+\ii s\bm\alpha_\delta,\bm\alpha_{\mathrm c}-\ii s\bm t_\delta)\) for every \(s\) in the disc used for the Taylor expansion. Then
\be
\mathcal F(\tau_A)=\left.e^{\ii\cD_\perp}\mathcal F(\rho_A(\bm t,\bm\alpha))\right|_{\bm t=\bm t_{\mathrm c},\,\bm\alpha=\bm\alpha_{\mathrm c}}.\label{eq-matrix-function-taylor-series}
\ee
If \(\mathcal F\) is defined by contour integration as in \cref{app-holomorphy-and-taylor-series}, the same contours and logarithm branch must be valid throughout this domain.

Taking \(\mathcal F=S\) in \cref{eq-matrix-function-taylor-series} gives the entropy equality in \cref{eq-transition-matrix-and-entropy-taylor-series}. \Cref{eq-real-imaginary-taylor-series} with \(\cD_{\bm y}=\cD_\perp\) gives the real part from the cosine series and the imaginary part from the sine series, both applied to ordinary entanglement entropy at the real midpoint.

\paragraph{Sufficiently close parameter values.} The following argument proves convergence of the entropy Taylor series for sufficiently close states. Let \((\bm t_0,\bm\alpha_0)\) be real parameter values of this family where all eigenvalues of \(\rho_A\) are positive. The normalisation in \cref{eq-holomorphic-matrix-family} is also positive there. By continuity, there is a complex neighbourhood in which the normalisation is nonzero and the spectrum stays in the right half-plane. The matrix and its entropy are holomorphic there by \cref{app-holomorphy-and-taylor-series}. For both real parameter values sufficiently close to \((\bm t_0,\bm\alpha_0)\), the values \((\bm t_{\mathrm c}+\ii s\bm\alpha_\delta,\bm\alpha_{\mathrm c}-\ii s\bm t_\delta)\) remain in this neighbourhood for every complex \(s\) with \(|s|<2\). The matrix and entropy Taylor series therefore converge at \(s=1\), which proves \cref{eq-transition-matrix-and-entropy-taylor-series} for all such pairs. The same argument applies to \(s\mapsto S(\bm t_{\mathrm c}+s\bm t_\delta,\bm\alpha_{\mathrm c}+s\bm\alpha_\delta)\), whose Taylor series then converges at \(s=\pm1\).

\paragraph{Ordered products of exponentials.} Let \(\ket{\psi(\bm t,\bm\alpha)}=\prod_{a=1}^de^{-(\alpha^a+\ii t^a)H_a}\ket{\Psi}\), with fixed matrices \(H_a\), a fixed state \(\ket{\Psi}\), and the factors in the displayed order, without assuming \([H_a,H_b]=0\). For two states in this family, \cref{eq-transition-matrix-and-entropy-taylor-series} gives their reduced transition matrix and pseudo entropy under the stated assumptions for the Taylor expansion. For one Hermitian Hamiltonian \(H\), choose \(\ket{\psi_1(t)}=e^{-\ii tH}\ket{\Psi}\) and \(\ket{\psi_2}=\ket{\Psi}\). Then the first-order result for pseudo entropy in \cite{Misumi:2026modular} follows by expanding the full right-hand side of \cref{eq-transition-matrix-and-entropy-taylor-series} about \(t=0\), including its \(t\)-dependent midpoint. The matrix identity and convergent Taylor series for the entropy apply to all orders without assuming a diagonal form. For each fixed positive integer \(n\), the matrix identity in \cref{eq-transition-matrix-complex-parameters} also extends to the \(n\)-th power and trace for trace-class operator families holomorphic in trace norm, as shown in \cref{app-holomorphy-and-taylor-series}.

\subsection{Average entanglement entropy, excess pseudo entropy, and R\'enyi entropies}
For \(i=1,2\), let \(S_i^{\mathrm E}\) be the entanglement entropy of \(\ket{\psi_i}\), and let \(S_{\mathrm{av}}^{\mathrm E}\) be their average. Define
\be
S_i^{\mathrm E}\coloneqq S^{\mathrm E}(\bm t_i,\bm\alpha_i),\qquad S_{\mathrm{av}}^{\mathrm E}\coloneqq\frac{S_1^{\mathrm E}+S_2^{\mathrm E}}{2},\qquad\cD_\parallel\coloneqq\sum_{a=1}^d\left(t_\delta^a\partial_{t^a}+\alpha_\delta^a\partial_{\alpha^a}\right).
\ee
In the following formulae, \(e^{\pm\cD_\parallel}S^{\mathrm E}(\bm t,\bm\alpha)\coloneqq S^{\mathrm E}(\bm t\pm\bm t_\delta,\bm\alpha\pm\bm\alpha_\delta)\), while \(\cosh(\cD_\parallel)\) and \(\sinh(\cD_\parallel)\) denote the half-sum and half-difference of these two values. These definitions do not require a Taylor series. Their representations as series of derivatives at the midpoint require separately that \(s\mapsto S(\bm t_{\mathrm c}+s\bm t_\delta,\bm\alpha_{\mathrm c}+s\bm\alpha_\delta)\) be holomorphic near \(-1\leq s\leq1\), with its Taylor series about \(s=0\) converging at \(s=\pm1\). This condition is independent of the one for \(\cD_\perp\). Evaluating at the two real parameter values gives
\be
S_2^{\mathrm E}=\left.e^{\,\cD_\parallel}S^{\mathrm E}(\bm t,\bm\alpha)\right|_{\bm t=\bm t_{\mathrm c},\,\bm\alpha=\bm\alpha_{\mathrm c}},\qquad S_1^{\mathrm E}=\left.e^{-\cD_\parallel}S^{\mathrm E}(\bm t,\bm\alpha)\right|_{\bm t=\bm t_{\mathrm c},\,\bm\alpha=\bm\alpha_{\mathrm c}}.\label{eq-entropies-at-two-parameter-values}
\ee
Adding and subtracting these relations gives
\be
S_{\mathrm{av}}^{\mathrm E}=\left.\cosh(\cD_\parallel)S^{\mathrm E}(\bm t,\bm\alpha)\right|_{\bm t=\bm t_{\mathrm c},\,\bm\alpha=\bm\alpha_{\mathrm c}},\qquad\frac{S_2^{\mathrm E}-S_1^{\mathrm E}}{2}=\left.\sinh(\cD_\parallel)S^{\mathrm E}(\bm t,\bm\alpha)\right|_{\bm t=\bm t_{\mathrm c},\,\bm\alpha=\bm\alpha_{\mathrm c}}.\label{eq-entropy-average-and-half-difference}
\ee
The excess compares the real part of pseudo entropy for the pair with the average entanglement entropy of the two states. Its sign was studied in examples where the two states are in the same or different quantum phases \cite{Mollabashi:2020GaussianPE}. The excess was also used to compare states associated with links in Chern-Simons theory \cite{Caputa:2024SVD}. Define the excess pseudo entropy by
\be
\Delta S_{12}\coloneqq\SR(\tau_A)-S_{\mathrm{av}}^{\mathrm E}=\left.\left[\cos(\cD_\perp)-\cosh(\cD_\parallel)\right]S^{\mathrm E}(\bm t,\bm\alpha)\right|_{\bm t=\bm t_{\mathrm c},\,\bm\alpha=\bm\alpha_{\mathrm c}}.\label{eq-excess-pseudo-entropy}
\ee
For sufficiently small \((\bm t_\delta,\bm\alpha_\delta)\) at a fixed midpoint \((\bm t_{\mathrm c},\bm\alpha_{\mathrm c})\) where \(S\) is holomorphic, expanding the cosine and hyperbolic cosine gives
\be
\Delta S_{12}=-\frac12\left.\left(\cD_\perp^{\,2}+\cD_\parallel^{\,2}\right)S^{\mathrm E}(\bm t,\bm\alpha)\right|_{\bm t=\bm t_{\mathrm c},\,\bm\alpha=\bm\alpha_{\mathrm c}}+\mathcal O\!\left(\left[\|\bm t_\delta\|^2+\|\bm\alpha_\delta\|^2\right]^2\right).\label{eq-excess-quadratic-term}
\ee
For \(d=1\), this reduces to
\be
\Delta S_{12}=-\frac12\left(t_\delta^2+\alpha_\delta^2\right)\left.\left(\partial_t^2+\partial_\alpha^2\right)S^{\mathrm E}(t,\alpha)\right|_{t=t_{\mathrm c},\,\alpha=\alpha_{\mathrm c}}+\mathcal O\!\left[\left(t_\delta^2+\alpha_\delta^2\right)^2\right].\label{eq-excess-quadratic-term-one-parameter-pair}
\ee
Thus \(\Delta S_{12}\) has no term linear in \((\bm t_\delta,\bm\alpha_\delta)\), and \cref{eq-excess-quadratic-term} gives its quadratic term.
\paragraph{R\'enyi entropies.} For every integer \(n\geq2\), \(\Tr[\rho_A(\bm t,\bm\alpha)^n]\) is a polynomial in the entries of the reduced matrix. Applying \cref{eq-entropy-taylor-series} to this polynomial gives
\be
\Tr\!\left[\left(\tau_A\right)^n\right]=\left.e^{\,\ii\cD_\perp}\Tr\!\left[\rho_A(\bm t,\bm\alpha)^n\right]\right|_{\bm t=\bm t_{\mathrm c},\,\bm\alpha=\bm\alpha_{\mathrm c}}.
\ee
For \(w\in(0,\infty)\setminus\{1\}\), with powers for \(w\notin\mathbb Z\) as in \cref{app-holomorphy-and-taylor-series}, the pseudo-R\'enyi and ordinary R\'enyi entropies are defined, respectively, by \cite{Nakata:2020luh,GuoZhang:2023sumrule}
\be
S_w(\tau_A)\coloneqq\frac{1}{1-w}\Log\Tr[\tau_A^w],\qquad S_w^{\mathrm E}(\bm t,\bm\alpha)\coloneqq\frac{1}{1-w}\log\Tr[\rho_A(\bm t,\bm\alpha)^w].
\ee
If \(\Tr[\rho_A(\bm t_{\mathrm c}+\ii s\bm\alpha_\delta,\bm\alpha_{\mathrm c}-\ii s\bm t_\delta)^n]\) is nonzero and remains in \(\mathbb C\setminus(-\infty,0]\) for every \(s\) in the disc \(|s|<r_s\), \(r_s>1\), used in the Taylor expansion, then
\be
S_n(\tau_A)=\left.e^{\,\ii\cD_\perp}S_n^{\mathrm E}(\bm t,\bm\alpha)\right|_{\bm t=\bm t_{\mathrm c},\,\bm\alpha=\bm\alpha_{\mathrm c}}.\label{eq-pseudo-renyi-continuation}
\ee
The first-order result in \cite{Misumi:2026bfk} follows from this. The formula for \(\Tr[(\tau_A)^n]\) requires no matrix logarithm and remains valid for singular reduced matrices. Since \(S_n^{\mathrm E}\) is real for real parameters, \cref{eq-real-imaginary-taylor-series} gives the real and imaginary parts with \(S\) and \(S^{\mathrm E}\) replaced by \(S_n\) and \(S_n^{\mathrm E}\). With \(S\) and \(S^{\mathrm E}\) replaced by \(S_n\) and \(S_n^{\mathrm E}\), \cref{eq-entropies-at-two-parameter-values,eq-entropy-average-and-half-difference} give the two ordinary entropies, their average and half-difference, and \cref{eq-excess-pseudo-entropy,eq-excess-quadratic-term,eq-excess-quadratic-term-one-parameter-pair} give the excess and its quadratic term. Here \(e^{\pm\cD_\parallel}S_n^{\mathrm E}(\bm t,\bm\alpha)\coloneqq S_n^{\mathrm E}(\bm t\pm\bm t_\delta,\bm\alpha\pm\bm\alpha_\delta)\), with the same half-sum and half-difference definitions. For their representations as series of derivatives at the midpoint, require the analytic continuation of \(s\mapsto S_n^{\mathrm E}(\bm t_{\mathrm c}+s\bm t_\delta,\bm\alpha_{\mathrm c}+s\bm\alpha_\delta)\), with the same logarithm branch, to be holomorphic near \(-1\leq s\leq1\), with its Taylor series about \(s=0\) converging at \(s=\pm1\).

The extension to complex R\'enyi order \(w\) is given in \cref{app-holomorphy-and-taylor-series}.

\subsection{Three density matrices}
The reduced transition matrix can be obtained from three reduced density matrices and their normalisations, as in \cite{GuoZhang:2023sumrule}. One can rescale the states without changing \(\tau_A\) so that \(\braket{\psi_i}{\psi_i}=1\), \(i=1,2\). For real \(u\), define
\be
\begin{aligned} &\ket{\varphi_{\mathrm c}}\coloneqq\frac{\ket{\psi_1}+\ket{\psi_2}}2, & \quad&\ket{\varphi_\delta}\coloneqq\frac{\ket{\psi_2}-\ket{\psi_1}}2, & \quad&\ket{\varphi(u)}\coloneqq\ket{\varphi_{\mathrm c}}+\ii u\ket{\varphi_\delta}, \\ &\mathcal N(u)\coloneqq\braket{\varphi(u)}{\varphi(u)}, & \quad&\rho_A(u)\coloneqq\Tr_B\frac{\proj{\varphi(u)}}{\mathcal N(u)}. \end{aligned}\label{eq-superposition-state-family}
\ee
One obtains \(\ket{\varphi(\ii)}=\ket{\psi_1}\) and \(\ket{\varphi(-\ii)}=\ket{\psi_2}\). The unnormalised reduced matrix \(\mathcal N(u)\rho_A(u)\) is a polynomial of degree at most two in real \(u\). Continuing this polynomial to \(u=\ii\) gives \(\braket{\psi_2}{\psi_1}\tau_A\), and its trace gives \(\braket{\psi_2}{\psi_1}\). For \(\mathcal N(-1),\mathcal N(0),\mathcal N(1)>0\), interpolation from these three values therefore gives
\be
\tau_A=\frac{2\mathcal N(0)\rho_A(0)-\frac{1+\ii}{2}\mathcal N(-1)\rho_A(-1)-\frac{1-\ii}{2}\mathcal N(1)\rho_A(1)}{2\mathcal N(0)-\frac{1+\ii}{2}\mathcal N(-1)-\frac{1-\ii}{2}\mathcal N(1)}.\label{eq-three-density-matrix-formula}
\ee
The denominator is \(\braket{\psi_2}{\psi_1}\), so the formula gives \(\tau_A\) from the three normalisation factors and reduced density matrices, not from three entanglement entropies. \Cref{app-superposition-sum-rule} shows that the formula in \cite{GuoZhang:2023sumrule} evaluates the same polynomial \(\mathcal N(u)\rho_A(u)\) at \(u=\ii\), using \(u=0,\pm\sqrt3\) rather than \(u=-1,0,1\). With the change of variables and weights given there, the comparison also applies to \([\mathcal N(u)\rho_A(u)]^n\) for every fixed integer \(n\geq1\).
\paragraph{Interpolation at real parameter values.} If \(\ket{\psi_2}\) is a nonzero scalar multiple of \(\ket{\psi_1}\), one ordinary reduced density matrix already gives \(\tau_A\). If no nonzero scalar relates the two states, \(2n+1\) real values give \(\Tr[(\tau_A)^n]\) for the family in \cref{eq-superposition-state-family}. More generally, for a state family whose coefficients have degree at most \(r\), \(2r+1\) real values give the reduced transition matrix and \(2nr+1\) give \(\Tr[(\tau_A)^n]\), as shown in \cref{app-polynomial-interpolation}.

\subsection{Two-qubit example}
We compare the real part of pseudo entropy with the average entanglement entropy and compute the imaginary part from both the reduced transition matrix and the Cauchy-Riemann equations.

Consider two qubits with
\be
\sigma_x\coloneqq\left(\begin{smallmatrix}0&1\\1&0\end{smallmatrix}\right),\qquad H\coloneqq\sigma_x\otimes\sigma_x,\qquad\ket{\Psi}\coloneqq\ket{00}.
\ee
Since \(H^2=\one\), \(e^{-(\alpha+\ii t)H}=\cosh(\alpha+\ii t)\one-\sinh(\alpha+\ii t)H\), so after normalisation
\be
\ket{\psi(t,\alpha)}=\frac{\cosh(\alpha+\ii t)\ket{00}-\sinh(\alpha+\ii t)\ket{11}}{\sqrt{\cosh(2\alpha)}},\qquad t,\alpha\in\mathbb R.\label{eq-qubit-state}
\ee
This is the \(d=1\) example of the family \(\ket{\psi(t,\alpha)}=e^{-(\alpha+\ii t)H}\ket{\Psi}\) considered in \cref{sec-two-states}, so \cref{eq-transition-matrix-complex-parameters} applies. Tracing out either qubit gives
\be
\rho_A(t,\alpha)=\frac12\operatorname{diag}\!\left(1+\frac{\cos(2t)}{\cosh(2\alpha)},1-\frac{\cos(2t)}{\cosh(2\alpha)}\right).
\ee

\begin{figure}[h!]
\centering\includegraphics[width=\textwidth]{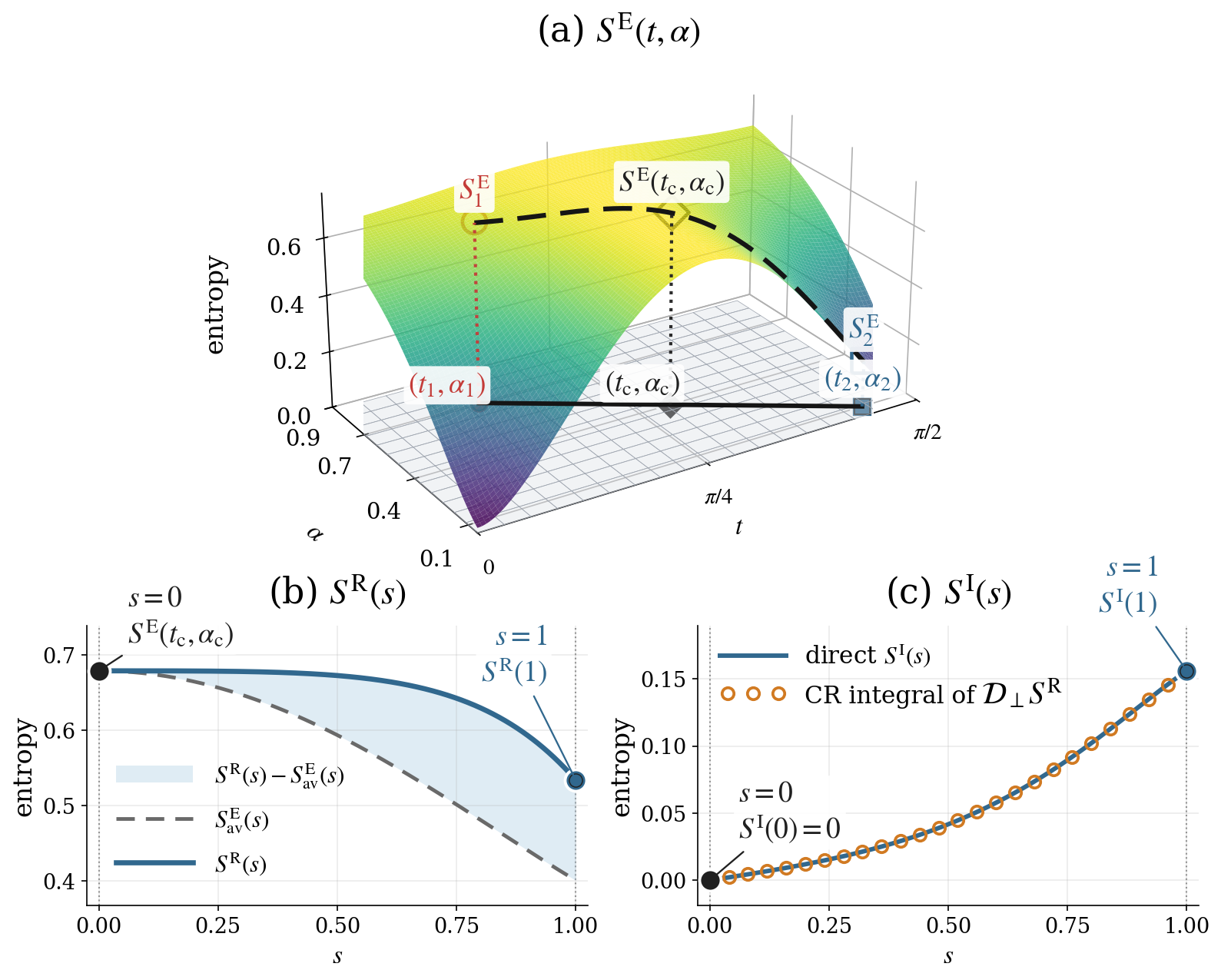} \caption{Two-qubit example. (a) The entanglement entropy for \cref{eq-qubit-state}. The solid line is the segment joining the two chosen points in the \((t,\alpha)\) plane, and the dashed curve is the corresponding entropy along that segment. The grey plane only shows the \((t,\alpha)\) coordinates and does not represent zero entropy. (b) The real part \(\SR(s)\), the average \(S_{\mathrm{av}}^{\mathrm E}(s)\), and the excess pseudo entropy \(\SR(s)-S_{\mathrm{av}}^{\mathrm E}(s)\). (c) The imaginary part computed from the eigenvalues of the reduced transition matrix, compared with the result of integrating the Cauchy-Riemann equation in \cref{eq-qubit-imaginary-part-integral}. In (b) and (c), \(s=0\) gives the same state at the midpoint, while \(s=1\) gives the two chosen states in \cref{eq-qubit-parameters}.}\label{fig-qubit-example}
\end{figure}
For \cref{fig-qubit-example}, choose the two parameter pairs
\be
(t_1,\alpha_1)=(0.4,0.7),\quad(t_2,\alpha_2)=(1.4,0.1)\;\Rightarrow\;(t_{\mathrm c},\alpha_{\mathrm c})=(0.9,0.4),\quad(t_\delta,\alpha_\delta)=(0.5,-0.3).\label{eq-qubit-parameters}
\ee
For \(0\leq s\leq1\), let the two parameter pairs depend on \(s\) by
\be
t_{1,2}(s)=t_{\mathrm c}\mp st_\delta,\qquad\alpha_{1,2}(s)=\alpha_{\mathrm c}\mp s\alpha_\delta,\qquad\tau_A(s)=\rho_A(t_{\mathrm c}+\ii s\alpha_\delta,\alpha_{\mathrm c}-\ii st_\delta).\label{eq-qubit-path}
\ee
The last equality in \cref{eq-qubit-path} is \cref{eq-transition-matrix-complex-parameters} with \((t_\delta,\alpha_\delta)\) replaced by \(s(t_\delta,\alpha_\delta)\). At \(s=0\) the two states coincide at \((t_{\mathrm c},\alpha_{\mathrm c})\), while \(s=1\) gives the two states in \cref{eq-qubit-parameters}. Write \(S(\tau_A(s))=\SR(s)+\ii\SI(s)\). By \cref{eq-transition-matrix-and-entropy-taylor-series,eq-real-imaginary-taylor-series}, \(\SR(s)=\left.\cos(s\cD_\perp)S^{\mathrm E}(t,\alpha)\right|_{t=t_{\mathrm c},\,\alpha=\alpha_{\mathrm c}}\). The average of the two entanglement entropies is
\be
S_{\mathrm{av}}^{\mathrm E}(s)\coloneqq\frac12\left[S^{\mathrm E}(t_1(s),\alpha_1(s))+S^{\mathrm E}(t_2(s),\alpha_2(s))\right],
\ee
so, by \cref{eq-excess-pseudo-entropy}, \(\SR(s)-S_{\mathrm{av}}^{\mathrm E}(s)\) is the excess pseudo entropy. For the parameters in \cref{eq-qubit-parameters}, \(\cD_\perp=\alpha_\delta\partial_t-t_\delta\partial_\alpha=-0.3\,\partial_t-0.5\,\partial_\alpha\) by \cref{eq-D-perp-definition}, and the two eigenvalues remain outside \((-\infty,0]\) for \(0\leq s\leq1\). For the parameter directions in \(\cD_\perp\) and \(\cD_\parallel\), all zeros and poles of the eigenvalues have \(|s|\geq\sqrt{0.4^2+(\pi/2-0.9)^2}/\sqrt{0.3^2+0.5^2}>1\). The entropy continued from \(s=0\) is therefore holomorphic on a disc \(|s|<r_s\) with \(r_s>1\), so the entropy Taylor series converge at \(s=\pm1\). Since \(\SI(0)=0\), integrating the Cauchy-Riemann relations in \cref{eq-cauchy-riemann-equations} along \cref{eq-qubit-path} gives
\be
\SI(s)=\int_0^s\left.\cD_\perp\SR\!\left(\rho_A(t+\ii s'\alpha_\delta,\alpha-\ii s't_\delta)\right)\right|_{t=t_{\mathrm c},\,\alpha=\alpha_{\mathrm c}}\dd s'.\label{eq-qubit-imaginary-part-integral}
\ee
The real part in the integrand is evaluated at complex parameter values along \cref{eq-qubit-path}.

For the same family at \(\alpha=0\), \(\rho_A(t,0)=\operatorname{diag}(\cos^2t,\sin^2t)\). For real \(t\to0\), \(S^{\mathrm E}(t,0)=-t^2\log(t^2)+\mathcal O(t^2)\). The density matrix is holomorphic at \(t=0\), but its entropy is not.
\FloatBarrier

\section{Conformal field theory}\label{sec-cft}
We consider a two-dimensional CFT on the infinite line, with Hamiltonian \(H\) and conformal boundary state \(\ket{\mathcal B}\). The subsystem \(A=[0,\ell]\) is an interval of length \(\ell\), and \(B\) is its complement. The parameter \(\alpha>0\) gives Euclidean evolution that regularises the boundary state. \(t\) is real time and \(\eps>0\) is the ultraviolet cutoff.

For equal times \(t_1=t_2=t\) and different \(\alpha_1,\alpha_2\), the state formula below gives the shift \(t\mapsto t+\ii\alpha_\delta\), as in \cref{eq-transition-matrix-complex-parameters}, with the parameters defined in \cref{eq-parameter-centre-and-half-difference}. The explicit entropy and integral bound are for the free Dirac CFT. For states at opposite times, Kramers-Kronig relations give the first moment of the imaginary part. We evaluate it for the free Dirac CFT and, under the stated assumptions, for general conformal boundary states and the thermofield double.

For the boundary-state calculations below, \(\eps>0\), \(\alpha_i>0\), and \(\eps\ll\min(\ell,\alpha_i)\). The density and transition matrices are first defined at finite spatial volume and ultraviolet cutoff, with nonzero normalisations. The infinite-line formulae are limits of these normalised reduced expressions for a fixed interval. The normalisation and partial trace are taken before the limit. Whenever a matrix identity or analytic continuation of the entropy is passed to this limit, we assume locally uniform convergence on the complex parameter domain. Infinite-line limits are taken before large-time limits and time integrals. These regulator conventions also apply to the CFT appendix.

For real parameters, \(S^{\mathrm E}(t,\alpha)\) denotes the entanglement entropy in \cref{eq-CFT-entanglement-entropy}, with analytic continuation \(S(t,\alpha)\). In this section and the CFT appendix, \(\log\) at complex arguments denotes the logarithm branch fixed by continuation from real parameters, while \(\Log\) retains its principal-branch meaning. The symbols \(S\), \(\SR\), and \(\SI\), including when their argument is \(\tau_A\), denote this analytic continuation and its real and imaginary parts. The principal matrix logarithm in \cref{sec-entropy-continuation} gives the same entropy wherever it extends holomorphically from the same real parameter values on a connected domain. Evaluating the principal matrix logarithm only at the final parameter value can give a different branch. Taylor series are used only when the chosen continuation is holomorphic near the relevant parameter segment and the series converges at the required value.

A global quench in a CFT can be described by Euclidean evolution of a boundary state \cite{CalabreseCardy:2005Quench,Mollabashi:2021xsd}. The state below has the form \(e^{-(\alpha+\ii t)H}\ket{\Psi}\) considered in \cref{sec-two-states}, with \(\ket{\Psi}=\ket{\mathcal B}\). Using the regularisation and limits stated above, define
\be
\ket{\psi(t,\alpha)}\coloneqq e^{-\ii tH}e^{-\alpha H}\ket{\mathcal B},\qquad\alpha>0,\qquad\rho(t,\alpha)\coloneqq\frac{e^{-\ii tH}e^{-\alpha H}\proj{\mathcal B}e^{-\alpha H}e^{+\ii tH}}{\bra{\mathcal B}e^{-2\alpha H}\ket{\mathcal B}}.\label{eq-CFT-density-matrix}
\ee
Set \(\rho_A(t,\alpha)\coloneqq\Tr_B\rho(t,\alpha)\). At complex parameters, \(\rho(t,\alpha)\) denotes the analytic continuation of the expression in \cref{eq-CFT-density-matrix}. For the free Dirac fermion CFT with central charge \(c=1\) considered in \cite{Casini:2009sr,Mollabashi:2021xsd}, the calculation gives, for real \(t,\alpha\),
\be
S^{\mathrm E}(t,\alpha)=\frac16\log\!\left[\left(\frac{4\alpha}{\pi\eps}\right)^2\frac{\cosh^2\!\left(\frac{\pi t}{2\alpha}\right)\sinh^2\!\left(\frac{\pi\ell}{4\alpha}\right)}{\cosh\!\left[\frac{\pi}{2\alpha}(t-\ell/2)\right]\cosh\!\left[\frac{\pi}{2\alpha}(t+\ell/2)\right]}\right].\label{eq-CFT-entanglement-entropy}
\ee
For the two states \(\ket{\psi_i}\coloneqq\ket{\psi(t_i,\alpha_i)}\), with real \(t_i\), \(\alpha_i>0\), and \(i=1,2\), define \(t_{\mathrm c},t_\delta,\alpha_{\mathrm c},\alpha_\delta\) as in \cref{eq-parameter-centre-and-half-difference}. Substituting \((t,\alpha)=(t_{\mathrm c}+\ii\alpha_\delta,\alpha_{\mathrm c}-\ii t_\delta)\) in \cref{eq-CFT-density-matrix} gives \(\ket{\psi_1}\bra{\psi_2}\) in the numerator and \(\braket{\psi_2}{\psi_1}\) in the denominator. Therefore,
\be
\tau=\rho(t_{\mathrm c}+\ii\alpha_\delta,\,\alpha_{\mathrm c}-\ii t_\delta),\qquad S(\tau_A)=S(t_{\mathrm c}+\ii\alpha_\delta,\,\alpha_{\mathrm c}-\ii t_\delta).\label{eq-CFT-transition-matrix-complex-parameters}
\ee
The matrix equality holds for the conformal boundary states considered here. For the free Dirac CFT, the entropy equality follows separately from the regulated expression in \cref{eq-CFT-entanglement-entropy} under the conditions stated at the beginning of this section.

\subsection{Equal times and different values of \texorpdfstring{\(\alpha\)}{alpha}}
For \(t_1=t_2=t\), one has \(t_{\mathrm c}=t\), \(t_\delta=0\), and
\be
\tau_A(t)=\rho_A(t+\ii\alpha_\delta,\alpha_{\mathrm c}).\label{eq-CFT-equal-time-complex-parameter}
\ee
For equal real times and different \(\alpha_1,\alpha_2\), the calculation using a conformal map and twist operators in \cite{Mollabashi:2021xsd} gives the shift \(t\mapsto t+\ii\alpha_\delta\) and the regulated expression for \(\Tr[\tau_A(t)^n]\) for each fixed integer \(n\). \Cref{eq-CFT-equal-time-complex-parameter} is an equality of reduced matrices before taking a power, trace, or entropy. \Cref{app-cft-free-dirac} proves that the calculation using the conformal map and the matrix equality give the same moments of integer order and entropy. When a derivative with respect to \(w\) is used, a holomorphic continuation near \(n=1\) and locally uniform convergence jointly in that neighbourhood and the complex parameters are required. Since \(|\alpha_\delta|<\alpha_{\mathrm c}\), the Taylor series about real \(t\) converges at \(t+\ii\alpha_\delta\), as shown in \cref{app-cft-free-dirac}. Therefore,
\be
S(\tau_A(t))=e^{\,\ii\alpha_\delta\partial_t}S^{\mathrm E}(t,\alpha_{\mathrm c})=S(t+\ii\alpha_\delta,\alpha_{\mathrm c}).\label{eq-CFT-equal-time-entropy}
\ee
At fixed \(t\) and \(\alpha_{\mathrm c}\), \cref{eq-real-imaginary-taylor-series} gives the real and imaginary parts, including
\be
\SI(\tau_A(t))=\alpha_\delta\,\partial_tS^{\mathrm E}(t,\alpha_{\mathrm c})+\mathcal O(\alpha_\delta^3).\label{eq-CFT-imaginary-linear-term}
\ee
The average, half-difference, and excess follow from \cref{eq-entropy-average-and-half-difference,eq-excess-pseudo-entropy} with \(\cD_\perp=\alpha_\delta\partial_t\), \(\cD_\parallel=\alpha_\delta\partial_\alpha\), and evaluation at \((t,\alpha_{\mathrm c})\). Both entropy series converge for \(|\alpha_\delta|<\alpha_{\mathrm c}\), as shown in \cref{app-cft-free-dirac}.

\paragraph{Integral of the imaginary part.} Let \(S^{\mathrm E}(\infty,\alpha_{\mathrm c})\coloneqq\lim_{t\to\infty}S^{\mathrm E}(t,\alpha_{\mathrm c})\). The integral over real time relates the imaginary part to the change in ordinary entanglement entropy. At small \(\alpha_\delta\), its first-order term is \(\alpha_\delta[S^{\mathrm E}(\infty,\alpha_{\mathrm c})-S^{\mathrm E}(0,\alpha_{\mathrm c})]\), as derived in \cref{app-cft-free-dirac}. The real part has the limit \(S^{\mathrm E}(\infty,\alpha_{\mathrm c})\), so its time integral requires subtracting this constant to converge. Combining the formula derived in \cref{app-integral-formulae} with \cref{eq-CFT-entanglement-entropy} gives
\be
\int_0^\infty\SI(\tau_A(t))\,\dd t=\int_0^{\alpha_\delta}\left[S^{\mathrm E}(\infty,\alpha_{\mathrm c})-\SR(\ii y,\alpha_{\mathrm c})\right]\dd y=\frac16\int_0^{\alpha_\delta}\log\!\left[1+\frac{\sinh^2\!\left(\frac{\pi\ell}{4\alpha_{\mathrm c}}\right)}{\cos^2\!\left(\frac{\pi y}{2\alpha_{\mathrm c}}\right)}\right]\dd y.\label{eq-CFT-imaginary-part-integral}
\ee
Because the integrand is positive and even in \(y\), the left-hand side of \cref{eq-CFT-imaginary-part-integral} is an odd, strictly increasing function of \(\alpha_\delta\) for \(|\alpha_\delta|<\alpha_{\mathrm c}\). Its limiting value, derived in \cref{app-cft-free-dirac}, gives
\be
\left|\int_0^\infty\SI(\tau_A(t))\,\dd t\right|<\frac{\pi\ell}{12},\qquad\lim_{\alpha_\delta\to\pm\alpha_{\mathrm c}}\int_0^\infty\SI(\tau_A(t))\,\dd t=\pm\frac{\pi\ell}{12}.\label{eq-CFT-imaginary-part-integral-bound}
\ee
At fixed \(\alpha_{\mathrm c}\), \(\alpha_\delta\to+\alpha_{\mathrm c}\) gives \((\alpha_1,\alpha_2)\to(0,2\alpha_{\mathrm c})\), while \(\alpha_\delta\to-\alpha_{\mathrm c}\) gives \((\alpha_1,\alpha_2)\to(2\alpha_{\mathrm c},0)\). These limiting values give the bound, independent of \(\alpha_{\mathrm c}\), while the absolute value of the integral remains below it for positive \(\alpha_1,\alpha_2\). These are limits of the continuum expression, with the ultraviolet limit taken first, or jointly so that \(\eps/\min(\ell,\alpha_1,\alpha_2)\to0\). The small-\(\alpha_\delta\) expansion is derived in \cref{app-cft-free-dirac}.

\subsection{States at opposite times and Kramers-Kronig relations}
For fixed \(\alpha\), write \(\zeta\coloneqq\alpha+\ii t\). Under the holomorphy and decay assumptions in \cref{app-integral-formulae}, the real and imaginary parts obey Kramers-Kronig relations, and the coefficient in the large-\(|\zeta|\) expansion gives the first moment of the imaginary part. Set \((t_1,\alpha_1)=(t,\alpha)\) and \((t_2,\alpha_2)=(-t,\alpha)\). Then
\begin{align}
&\tau(t,\alpha)=\frac{e^{-\zeta H}\proj{\mathcal B}e^{-\zeta H}}{\bra{\mathcal B}e^{-2\zeta H}\ket{\mathcal B}},\qquad\tau_A(t,\alpha)=\rho_A(0,\zeta),\\ &S(\tau_A(t,\alpha))=e^{\ii t\partial_\alpha}S^{\mathrm E}(0,\alpha)=S(0,\alpha+\ii t),\qquad|t|<\alpha.
\end{align}
Analytically continuing the \(t=0\) expression in \cref{eq-CFT-entanglement-entropy} from \(\alpha>0\) to complex \(\zeta\), define
\be
F(\zeta)\coloneqq S(0,\zeta)-\frac13\log(\ell/\eps)=\frac13\log\!\left[\frac{4\zeta}{\pi\ell}\tanh\!\left(\frac{\pi\ell}{4\zeta}\right)\right].\label{eq-CFT-vacuum-subtracted-entropy}
\ee
The function \(F\) is holomorphic for \(\operatorname{Re}\zeta>0\), so \(F(\alpha+\ii t)\) is defined for every real \(t\). The disc \(|\zeta-\alpha|<\alpha\) lies in the right half-plane, so the Taylor series about \(\zeta=\alpha\) converges for \(|t|<\alpha\). The singularity at \(\zeta=0\) implies that this Taylor series does not converge for \(|t|>\alpha\), although \(F(\alpha+\ii t)\) is defined there. For the free Dirac CFT, \(S_1^{\mathrm E}=S_2^{\mathrm E}=S^{\mathrm E}(t,\alpha)\). Thus the average is \(S^{\mathrm E}(t,\alpha)\), the half-difference is zero, and \(\Delta S_{12}=\SR(0,\alpha+\ii t)-S^{\mathrm E}(t,\alpha)\), as in \cref{eq-entropy-average-and-half-difference,eq-excess-pseudo-entropy}. Write \(F^{\mathrm R}\coloneqq\operatorname{Re}F\) and \(F^{\mathrm I}\coloneqq\operatorname{Im}F\). Sufficient conditions for the Kramers-Kronig relations and the first moment, as in \cref{app-integral-formulae}, are that \(F\) is holomorphic for \(\operatorname{Re}\zeta>\alpha_*\), real for real \(\zeta>\alpha_*\), and satisfies \(F(\zeta)=C_2/\zeta^2+\mathcal O(|\zeta|^{-2-\eta})\) as \(|\zeta|\to\infty\), uniformly for \(\operatorname{Re}\zeta\geq\alpha\), for every fixed \(\alpha>\alpha_*\geq0\), with \(C_2\in\mathbb R\) and \(\eta>0\). With \(\operatorname{PV}\) denoting the Cauchy principal value, the Kramers-Kronig relations give
\be
F^{\mathrm I}(\alpha+\ii t)=\frac1\pi\operatorname{PV}\int_{-\infty}^{\infty}\frac{F^{\mathrm R}(\alpha+\ii s)}{s-t}\,\dd s,\quad F^{\mathrm R}(\alpha+\ii t)=-\frac1\pi\operatorname{PV}\int_{-\infty}^{\infty}\frac{F^{\mathrm I}(\alpha+\ii s)}{s-t}\,\dd s.\label{eq-CFT-Kramers-Kronig}
\ee
Relations of this form for thermal pseudo entropy were derived in \cite{Caputa:2024thermal}.
\paragraph{First moment of the imaginary part.} \Cref{app-integral-formulae} derives the first moment of the imaginary part from the second relation in \cref{eq-CFT-Kramers-Kronig}, and \cref{app-cft-free-dirac} evaluates it for \(F\) in \cref{eq-CFT-vacuum-subtracted-entropy}. Since the vacuum subtraction is real, the free Dirac CFT gives
\be
\int_0^\infty t\,\SI(\tau_A(t,\alpha))\,\dd t=\frac{\pi^3\ell^2}{288}.\label{eq-free-dirac-first-moment}
\ee
\paragraph{Conformal boundary states.} For a unitary two-dimensional CFT on the infinite line with conformal boundary state \(\ket{\mathcal B}\), let \(\tau_{A,\mathcal B}\) be the reduced transition matrix and define
\be
\Sigma_{1,\mathcal B}\coloneqq\sum_{\Delta_a=1}\bigl(b_a^{(\mathcal B)}\bigr)^2,
\ee
where \(b_a^{(\mathcal B)}\) are the coefficients in the one-point functions of the spinless dimension-one bulk primaries in the normalisation specified in \cref{app-cft-short-interval-first-moment}. Assume that \(b_a^{(\mathcal B)}\neq0\) only when \(\Delta_a\geq1\), and that the analytically continued vacuum-subtracted entropy is holomorphic in a right half-plane as in \cref{app-integral-formulae}, with the uniform remainder bound for large \(|\zeta|\) stated in \cref{app-cft-short-interval-first-moment}. Then
\begin{align}
&-\frac{576}{\pi^3\ell^2} \int_0^\infty t\,\SI(\tau_{A,\mathcal B}(t,\alpha))\,\dd t = c-\frac32\Sigma_{1,\mathcal B},\label{eq-CFT-boundary-first-moment} \\ &\Sigma_{1,\mathcal B}=0 \quad\Longrightarrow\quad c = -\frac{576}{\pi^3\ell^2} \int_0^\infty t\,\SI(\tau_{A,\mathcal B}(t,\alpha))\,\dd t.\label{eq-CFT-central-charge-first-moment}
\end{align}
For the free Dirac example, \cref{eq-free-dirac-first-moment,eq-CFT-boundary-first-moment} with \(c=1\) give
\be
\Sigma_{1,\mathcal B}=2.
\ee
\Cref{app-free-dirac-dimension-one-coefficients} obtains \(\Sigma_{1,\mathcal B}=2\) independently from the boundary one-point functions for both the Neumann and Dirichlet families considered there. The function in \cref{eq-CFT-vacuum-subtracted-entropy} satisfies the assumptions on holomorphy and the uniform remainder bound for large \(|\zeta|\) used for \cref{eq-CFT-boundary-first-moment}. These remain assumptions for a general conformal boundary state. For these states, \cref{eq-real-imaginary-taylor-series} gives the real and imaginary parts, and \cref{eq-entropy-average-and-half-difference,eq-excess-pseudo-entropy} give the average, half-difference, and excess under the corresponding entropy and convergence conditions.
\paragraph{Thermofield double.} For each copy \(\mathrm L,\mathrm R\) of the same CFT, let \(\ket n\) be an energy eigenstate with energy \(E_n\). Use finite spatial volume as an infrared regulator and define
\be
\ket{\operatorname{TFD}(\beta,t)}\coloneqq\frac{1}{\sqrt{Z(\beta)}}\sum_ne^{-\beta E_n/2-\ii tE_n}\ket{n}_{\mathrm L}\ket{n}_{\mathrm R},\qquad Z(\beta)\coloneqq\sum_ne^{-\beta E_n},\qquad\beta>0.
\ee
For two such states with inverse temperatures \(\beta_1,\beta_2>0\) and times \(t_1,t_2\), define
\be
\beta_{\mathrm c}\coloneqq\frac{\beta_1+\beta_2}{2},\qquad s\coloneqq t_1-t_2,\qquad\zeta_\beta\coloneqq\beta_{\mathrm c}+\ii s.\label{eq-TFD-parameters}
\ee
Their reduced transition matrix on one copy is \(e^{-\zeta_\beta H}/Z(\zeta_\beta)\), the thermal density-matrix formula at the complex inverse temperature \(\zeta_\beta\) \cite{GotoNozakiTamaoka:2021,Caputa:2024thermal}. At finite spatial volume, the transition matrix requires \(Z(\beta_{\mathrm c}+\ii s)\neq0\) throughout the integration range. For an interval of length \(\ell\), let \(S(\zeta_\beta)\) denote its pseudo entropy, continued from positive real inverse temperature as specified above, in the infinite-line limit of \cref{app-cft-tfd}. The infinite-line limit is taken before the large-\(s\) limit and the integral below. Under the conditions in \cref{app-cft-tfd}, the first-moment formula in \cref{app-integral-formulae} gives
\be
c=-\frac{36}{\pi^3\ell^2}\int_0^\infty s\,\SI(\beta_{\mathrm c}+\ii s)\,\dd s.\label{eq-TFD-first-moment-central-charge}
\ee
In this limit, the integral is independent of \(\beta_{\mathrm c}\). For real \(\beta\), write \(S^{\mathrm E}(\beta)=S(\beta)\). The real and imaginary parts follow from \cref{eq-real-imaginary-taylor-series} with \(\cD_{\bm y}=\cD_\perp=s\partial_\beta\), evaluated at \(\beta=\beta_{\mathrm c}\), whenever the Taylor series converges. The average, half-difference, and excess follow from \cref{eq-entropy-average-and-half-difference,eq-excess-pseudo-entropy} with \(\cD_\parallel=(\beta_2-\beta_1)\partial_\beta/2\), under the corresponding logarithm and convergence conditions.

\section{Gaussian states and quenches}\label{sec-gaussian}
For pure Gaussian states of finitely many fermionic modes, the Hilbert space is finite-dimensional and \cref{sec-entropy-continuation} applies. The Fock space of bosonic Gaussian states is infinite-dimensional, so we use wavefunctions and covariance matrices. We show when the transition matrix for one oscillator or several independent modes equals an ordinary Gaussian density-matrix formula evaluated at complex time. For the free scalar lattice, interpolation of covariance matrices remains valid without the additional mass condition required by the complex-time formula. The last subsection compares the scalar and CFT calculations with Figure 21 of \cite{Mollabashi:2021xsd}.

\subsection{Pure Gaussian states of finitely many fermionic modes}
The representation given by Thouless' theorem has polynomial state coefficients. Therefore finitely many ordinary Gaussian states and their normalisations give the reduced transition matrix. The rank specified below gives the number of real parameter values in each interpolation formula.

Let \(\beta_i\), \(i=1,\ldots,N\), be fermionic annihilation operators, with creation operators \(\beta_i^\dagger\), and let \(\ket{\Phi_0}\) be their quasiparticle vacuum, satisfying \(\beta_i\ket{\Phi_0}=0\). Any quasiparticle vacuum nonorthogonal to \(\ket{\Phi_0}\) has the unnormalised form given by Thouless' theorem \cite{PorroDuguet:2022Onishi}
\be
\ket{\widetilde\Psi(\mathsf Z)}=\exp\!\left(\frac12\sum_{i,j=1}^{N}\beta_i^\dagger\mathsf Z_{ij}\beta_j^\dagger\right)\ket{\Phi_0},\qquad\mathsf Z^{\mathsf T}=-\mathsf Z.
\ee
The symmetric part of \(\mathsf Z\) gives zero in the exponent because \(\beta_i^\dagger\beta_j^\dagger=-\beta_j^\dagger\beta_i^\dagger\). Only the antisymmetric part is needed. The state coefficients in the occupation-number basis are polynomial in the independent entries of \(\mathsf Z\), and hence holomorphic. Any nonorthogonal pair of pure fermionic Gaussian states of the same parity can therefore be written in this form by choosing one state as the reference \cite{PorroDuguet:2022Onishi}.

The local argument in \cref{sec-two-states} proves convergence of the entropy Taylor series for sufficiently close states near real parameter values where all eigenvalues of the reduced density matrix are positive. For example, for two modes with \(\ket{\Phi_0}=\ket{00}\) and \((\mathsf Z)_{12}=a>0\), the state is \(\ket{00}+a\ket{11}\). After normalisation, tracing out either mode gives the positive eigenvalues \(1/(1+a^2)\) and \(a^2/(1+a^2)\).

For two such nonorthogonal states \(\ket{\widetilde\Psi_j}=\ket{\widetilde\Psi(\mathsf Z_j)}\), \(j=1,2\), let \(\tau_A\) be the reduced transition matrix and define
\be
\begin{aligned} &\mathsf Z_{\mathrm c}\coloneqq\frac{\mathsf Z_1+\mathsf Z_2}{2}, & \qquad&\mathsf Z_\delta\coloneqq\frac{\mathsf Z_2-\mathsf Z_1}{2}, & \qquad&\mathsf Z(u)\coloneqq\mathsf Z_{\mathrm c}+\ii u\mathsf Z_\delta, \\ &\ket{\widetilde\Psi(u)}\coloneqq\ket{\widetilde\Psi(\mathsf Z(u))}, & &\mathcal N(u)\coloneqq\braket{\widetilde\Psi(u)} {\widetilde\Psi(u)}, & &\rho_A(u)\coloneqq\Tr_B \frac{ \ket{\widetilde\Psi(u)} \bra{\widetilde\Psi(u)} }{ \mathcal N(u) }. \end{aligned}\label{eq-fermion-polynomial-family}
\ee
\paragraph{Interpolation at real values of \(u\).} Here \(\mathsf Z(\ii)=\mathsf Z_1\), \(\mathsf Z(-\ii)=\mathsf Z_2\). For real \(u\), the normalised state is an ordinary pure fermionic Gaussian state. Define the Lagrange interpolation polynomials by
\be
L_k^{(p)}(u)\coloneqq\prod_{\substack{j=0\\j\neq k}}^p\frac{u-u_j}{u_k-u_j},\qquad u_0,\ldots,u_p\in\mathbb R\ \text{distinct}.\label{eq-lagrange-polynomials}
\ee
If \(r\coloneqq\frac12\operatorname{rank}\mathsf Z_\delta\), the unnormalised reduced matrix \(\mathcal N(u)\rho_A(u)\) has exact degree \(2r\) in \(u\), as shown in \cref{app-fermion-interpolation}. Any \(2r+1\) distinct real values \(u_k\) give
\be
\tau_A=\frac{\displaystyle\sum_{k=0}^{2r}L_k^{(2r)}(\ii)\,\mathcal N(u_k)\,\rho_A(u_k)}{\displaystyle\sum_{k=0}^{2r}L_k^{(2r)}(\ii)\,\mathcal N(u_k)}.
\ee
The polynomial \(\Tr\!\left[\bigl(\mathcal N(u)\rho_A(u)\bigr)^n\right]\) has exact degree \(2nr\). Therefore \(\Tr[(\tau_A)^n]\) is given by \(\mathcal N(u_k)\) and \(\Tr[\rho_A(u_k)^n]\) at any \(2nr+1\) distinct real points. The change of variables in \cref{app-superposition-sum-rule} applies to the family in \cref{eq-superposition-state-family}, not to the polynomial dependence on \(\mathsf Z\) used here. Under the conditions of \cref{sec-entropy-continuation}, \cref{eq-transition-matrix-and-entropy-taylor-series,eq-pseudo-renyi-continuation} give the pseudo entropy and pseudo-R\'enyi entropy for each fixed \(n\). \Cref{app-fermion-interpolation} also expresses \(\langle\widetilde\Psi_2|(O_A\otimes\one_B)|\widetilde\Psi_1\rangle\), for any subsystem operator \(O_A\), through expectation values and normalisations of ordinary Gaussian states. Writing the independent entries of \(\mathsf Z\) as \(\bm\alpha+\ii\bm t\), \cref{eq-real-imaginary-taylor-series} with \(\cD_{\bm y}=\cD_\perp\) gives the real and imaginary parts, while \cref{eq-entropy-average-and-half-difference,eq-excess-pseudo-entropy} give the average, half-difference, and excess under the separate convergence conditions stated above.

\subsection{Bosonic Gaussian states}
Even finitely many bosonic modes have an infinite-dimensional Fock space, so the finite-dimensional argument in \cref{sec-entropy-continuation} does not apply. Let \(q_i\) and \(p_i\), \(i=1,\ldots,N\), be the canonical position and momentum operators, satisfying
\be
[q_i,p_j]=\ii\delta_{ij},\qquad[q_i,q_j]=[p_i,p_j]=0.
\ee
Let \(\ket{\bm q}\) denote their joint position eigenstate, with eigenvalue vector \(\bm q=(q_1,\ldots,q_N)^{\mathsf T}\). A zero-mean pure Gaussian state \(\ket{\Psi(\mathsf X,\mathsf Y)}\) has position-space wavefunction \cite{Menicucci:2011Gaussian}
\be
\langle\bm q|\Psi(\mathsf X,\mathsf Y)\rangle\propto\exp\!\left[-\frac12\bm q^{\mathsf T}(\mathsf X+\ii\mathsf Y)\bm q\right],\qquad\mathsf X=\mathsf X^{\mathsf T}>0,\qquad\mathsf Y=\mathsf Y^{\mathsf T},
\ee
where \(\mathsf X\) and \(\mathsf Y\) are real. With
\be
\bm\xi\coloneqq(q_1,\ldots,q_N,p_1,\ldots,p_N)^{\mathsf T},
\ee
where \(\xi_b\) denotes the \(b\)-th component of \(\bm\xi\), \(b=1,\ldots,2N\), its covariance matrix is defined by
\be
\Gamma_{bc}\coloneqq\frac{\langle\Psi(\mathsf X,\mathsf Y)|\{\xi_b,\xi_c\}|\Psi(\mathsf X,\mathsf Y)\rangle}{2\langle\Psi(\mathsf X,\mathsf Y)|\Psi(\mathsf X,\mathsf Y)\rangle},\qquad b,c=1,\ldots,2N,
\ee
and is
\be
\Gamma(\mathsf X,\mathsf Y) = \frac12 \begin{pmatrix} \mathsf X^{-1} & -\mathsf X^{-1}\mathsf Y \\ -\mathsf Y\mathsf X^{-1} & \mathsf X+\mathsf Y\mathsf X^{-1}\mathsf Y \end{pmatrix}.\label{eq-bosonic-covariance-matrix}
\ee
Choose subsystem \(A\) to contain the first \(N_A\) bosonic modes, with \(B\) its complement, and set
\be
\bm\xi_A \coloneqq(q_1,\ldots,q_{N_A},p_1,\ldots,p_{N_A})^{\mathsf T}, \qquad J_A \coloneqq\begin{pmatrix} 0&\one_{N_A}\\ -\one_{N_A}&0 \end{pmatrix}, \qquad[\xi_{A,b},\xi_{A,c}] = \ii(J_A)_{bc}.
\ee
For a Gaussian transition matrix \(\tau\), let \(\tau_A\coloneqq\Tr_B\tau\). Its covariance is \cite{Mollabashi:2021xsd}
\be
(\Gamma_A)_{bc} \coloneqq\frac12 \Tr[ \{\xi_{A,b},\xi_{A,c}\}\tau_A ], \qquad\Gamma_A = \begin{pmatrix} X&R\\ R^{\mathsf T}&P \end{pmatrix}.
\ee
\Cref{app-bosonic-gaussian-kernel-covariance} proves that the covariance formula remains valid at complex parameter values and that the reduced covariance is obtained by retaining the entries belonging to subsystem \(A\). For an ordinary Gaussian density matrix, the eigenvalues of \(\ii J_A\Gamma_A\) occur in pairs \(\pm\nu_1,\ldots,\pm\nu_{N_A}\), where the positive numbers \(\nu_j\) are the symplectic eigenvalues. At complex parameter values, \(\nu_1,\ldots,\nu_{N_A}\) denote the \(N_A\) eigenvalues obtained by continuing the positive symplectic eigenvalues as a group, counted with algebraic multiplicity. The conditions on spectral separation and contours used to continue \(S(\Gamma_A)\) holomorphically are given in \cref{app-bosonic-gaussian-kernel-covariance}. Define \cite{Mollabashi:2020GaussianPE,Mollabashi:2021xsd}
\be
S(\Gamma_A)\coloneqq\sum_{j=1}^{N_A}\left[\left(\nu_j+\frac12\right)\Log\!\left(\nu_j+\frac12\right)-\left(\nu_j-\frac12\right)\Log\!\left(\nu_j-\frac12\right)\right].\label{eq-bosonic-gaussian-entropy}
\ee
For an ordinary Gaussian density matrix, \cref{eq-bosonic-gaussian-entropy} is its von Neumann entropy \cite{Adesso:2014Gaussian}. At complex parameters, \(S(\Gamma_A)\) is defined by the contour formula in \cref{app-bosonic-gaussian-kernel-covariance}, without assuming equality to \(-\Tr[\tau_A\Log\tau_A]\) in the infinite-dimensional Fock space.
\paragraph{Two bosonic Gaussian states.} For real \(\mathsf X,\mathsf Y\), define
\be
\rho_A(\mathsf X,\mathsf Y)\coloneqq\Tr_B\frac{\ket{\Psi(\mathsf X,\mathsf Y)}\bra{\Psi(\mathsf X,\mathsf Y)}}{\braket{\Psi(\mathsf X,\mathsf Y)}{\Psi(\mathsf X,\mathsf Y)}},\qquad S^{\mathrm E}(\mathsf X,\mathsf Y)\coloneqq S\!\left(\Gamma_A(\mathsf X,\mathsf Y)\right).
\ee
The analytic continuations of \(\rho_A\), \(\Gamma_A\), and \(S(\Gamma_A)\) to complex \(\mathsf X,\mathsf Y\) are defined in \cref{app-bosonic-gaussian-kernel-covariance}. For two nonorthogonal states \(\ket{\Psi_j}=\ket{\Psi(\mathsf X_j,\mathsf Y_j)}\), \(j=1,2\), set
\be
\mathsf X_{\mathrm c}\coloneqq\frac{\mathsf X_1+\mathsf X_2}{2},\qquad\mathsf X_\delta\coloneqq\frac{\mathsf X_2-\mathsf X_1}{2},\qquad\mathsf Y_{\mathrm c}\coloneqq\frac{\mathsf Y_1+\mathsf Y_2}{2},\qquad\mathsf Y_\delta\coloneqq\frac{\mathsf Y_2-\mathsf Y_1}{2}.
\ee
Then
\begin{align}
&\mathsf X = \mathsf X_{\mathrm c}-\ii\mathsf Y_\delta, & &\mathsf Y = \mathsf Y_{\mathrm c}+\ii\mathsf X_\delta,\label{eq-bosonic-complex-parameters} \\ &\mathsf X+\ii\mathsf Y = \mathsf X_1+\ii\mathsf Y_1, & &\mathsf X-\ii\mathsf Y = \mathsf X_2-\ii\mathsf Y_2, \\ &\tau_A = \rho_A( \mathsf X_{\mathrm c}-\ii\mathsf Y_\delta,\, \mathsf Y_{\mathrm c}+\ii\mathsf X_\delta),\label{eq-bosonic-transition-matrix-complex-parameters} & &\Gamma_A^{1|2} = \Gamma_A( \mathsf X_{\mathrm c}-\ii\mathsf Y_\delta,\, \mathsf Y_{\mathrm c}+\ii\mathsf X_\delta).
\end{align}
Here \(\Gamma_A^{1|2}\) denotes the covariance of \(\tau_A\). \Cref{app-bosonic-gaussian-kernel-covariance} derives \(\tau_A\) and \(\Gamma_A^{1|2}\) from the Gaussian kernels. For the real symmetric matrices used in \cite{Mollabashi:2021xsd}, \cref{eq-bosonic-transition-matrix-complex-parameters} gives the Gaussian kernel of the transition matrix used there. Taking the partial trace gives its reduced kernel.

Since \(\mathsf X\) and \(\mathsf Y\) are symmetric, only their entries with \(a\leq b\) are independent. Define
\be
\begin{aligned} &\cD_\perp\coloneqq\sum_{1\leq a\leq b\leq N} \left[ (\mathsf X_\delta)_{ab} \frac{\partial}{\partial\mathsf Y_{ab}} - (\mathsf Y_\delta)_{ab} \frac{\partial}{\partial\mathsf X_{ab}} \right], \qquad S(\Gamma_A^{1|2}) = \left. e^{\,\ii\cD_\perp} S^{\mathrm E}(\mathsf X,\mathsf Y) \right|_{\mathsf X=\mathsf X_{\mathrm c},\, \mathsf Y=\mathsf Y_{\mathrm c}}, \\ &\cD_\parallel\coloneqq\sum_{1\leq a\leq b\leq N} \left[ (\mathsf X_\delta)_{ab} \frac{\partial}{\partial\mathsf X_{ab}} + (\mathsf Y_\delta)_{ab} \frac{\partial}{\partial\mathsf Y_{ab}} \right]. \end{aligned}\label{eq-bosonic-entropy-derivatives}
\ee
With \(S_i^{\mathrm E}\coloneqq S^{\mathrm E}(\mathsf X_i,\mathsf Y_i)\), \(i=1,2\), \(S_{\mathrm{av}}^{\mathrm E}\coloneqq(S_1^{\mathrm E}+S_2^{\mathrm E})/2\), and \(\Delta S_{12}\coloneqq\SR(\Gamma_A^{1|2})-S_{\mathrm{av}}^{\mathrm E}\), \cref{eq-real-imaginary-taylor-series} with \(\cD_{\bm y}=\cD_\perp\) gives the real and imaginary parts of the entropy in \cref{eq-bosonic-gaussian-entropy}. \Cref{eq-entropies-at-two-parameter-values,eq-entropy-average-and-half-difference,eq-excess-pseudo-entropy} give the two ordinary entropies, their average and half-difference, and the excess. All derivatives use \(\cD_\perp\) and \(\cD_\parallel\) defined above. Here \(e^{\pm\cD_\parallel}S^{\mathrm E}(\mathsf X,\mathsf Y)\coloneqq S^{\mathrm E}(\mathsf X\pm\mathsf X_\delta,\mathsf Y\pm\mathsf Y_\delta)\), and the result is evaluated at \((\mathsf X,\mathsf Y)=(\mathsf X_{\mathrm c},\mathsf Y_{\mathrm c})\). The hyperbolic cosine and sine give the half-sum and half-difference of these values. For each series of derivatives used, convergence and the conditions on spectral separation and contours of \cref{app-bosonic-gaussian-kernel-covariance} must be checked in its own parameter direction. The interpolation below gives \(\Gamma_A^{1|2}\) from ordinary covariance matrices and the known weights, after which \cref{eq-bosonic-gaussian-entropy} gives the entropy.

\paragraph{Sufficiently close parameter values.} The entropy Taylor series in both parameter directions converge for sufficiently close states near real parameter values where \(\mathsf X>0\) and \(\nu_j>1/2\) for every subsystem mode. By continuity, the Gaussian integral bounds in \cref{app-bosonic-gaussian-kernel-covariance} remain valid in a complex neighbourhood. The selected group of eigenvalues remains separated from the other eigenvalues and enclosed by one fixed contour, with the contour and its interior in \(\mathbb C\setminus(-\infty,1/2]\). The contour formula in \cref{app-bosonic-gaussian-kernel-covariance} then makes \(S(\Gamma_A)\) holomorphic there. As in \cref{sec-two-states}, the parameter values in each direction remain in this neighbourhood for every complex \(s\) with \(|s|<2\) when both real matrix pairs \((\mathsf X_i,\mathsf Y_i)\) are sufficiently close. The Taylor series for the entropy therefore converge at their required values. For example, two modes with \(\mathsf X_{11}=\mathsf X_{22}=2\), \(\mathsf X_{12}=\mathsf X_{21}=1\), \(\mathsf Y=0\), and \(A\) the first mode give \(\Gamma_A=\operatorname{diag}(1/3,1)\) and \(\nu_1=1/\sqrt3>1/2\).

\paragraph{Interpolation from ordinary Gaussian states.} Introduce a real parameter \(u\) and define
\begin{align}
&\mathsf X(u)\coloneqq\mathsf X_{\mathrm c}-u\mathsf Y_\delta, & &\mathsf Y(u)\coloneqq\mathsf Y_{\mathrm c}+u\mathsf X_\delta,\label{eq-bosonic-interpolation-family} \\ &r_{\mathrm b}\coloneqq\operatorname{rank}\mathsf Y_\delta, & &d_{\mathrm b}^{\max}\coloneqq\min(N+1,r_{\mathrm b}+2).
\end{align}
At \(u=\ii\), these are the complex parameters in \cref{eq-bosonic-complex-parameters}, while real \(u\) with \(\mathsf X(u)>0\) gives an ordinary Gaussian state. Let \(\Gamma_A(u)\) be the covariance of \(\rho_A(\mathsf X(u),\mathsf Y(u))\) whenever \(\mathsf X(u)>0\), and define
\be
\widetilde\Gamma_A(u)\coloneqq\det\mathsf X(u)\,\Gamma_A(u)=\sum_{k=0}^{d_{\mathrm b}^{\max}}\widetilde\Gamma_{A,k}u^k,\qquad d_{\mathrm b,A}\coloneqq\max\{k\mid\widetilde\Gamma_{A,k}\neq0\}.\label{eq-bosonic-covariance-degree}
\ee
Here \(\widetilde\Gamma_{A,k}\) is the coefficient of \(u^k\) in \(\widetilde\Gamma_A(u)\). The bound \(d_{\mathrm b,A}\leq d_{\mathrm b}^{\max}\) and the nonvanishing of \(\det\mathsf X(\ii)\) are proved in \cref{app-bosonic-covariance-polynomial-degree}. Since \(\mathsf X(0)=\mathsf X_{\mathrm c}>0\), one can choose \(d_{\mathrm b,A}+1\) distinct real values \(u_k\) sufficiently close to zero for which \(\mathsf X(u_k)>0\). Lagrange interpolation using \cref{eq-lagrange-polynomials} then gives
\be
\Gamma_A^{1|2}=\frac1{\det\mathsf X(\ii)}\sum_{k=0}^{d_{\mathrm b,A}}L_k^{(d_{\mathrm b,A})}(\ii)\,\det\mathsf X(u_k)\,\Gamma_A(u_k).\label{eq-bosonic-covariance-interpolation}
\ee
For the matrix family in \cref{eq-bosonic-interpolation-family}, \(d_{\mathrm b,A}+1\) ordinary reduced covariance matrices, with the known weights involving determinants in \cref{eq-bosonic-covariance-interpolation}, give \(\Gamma_A^{1|2}\). Before computing \(d_{\mathrm b,A}\), \(d_{\mathrm b}^{\max}+1\) is a sufficient number.

If \(\mathsf Y_1=\mathsf Y_2\), then \(\mathsf Y_\delta=0\) and \(\mathsf X(u)=\mathsf X_{\mathrm c}\). As shown in \cref{app-bosonic-covariance-polynomial-degree}, for subsystem \(A\) either \(d_{\mathrm b,A}=2\) or all \(u\)-dependent covariance coefficients vanish. In the first case, \(\Gamma_A(u)\) is quadratic in \(u\), so any three distinct real values give it. Since \(\mathsf X(u)=\mathsf X_{\mathrm c}>0\), every real value gives an ordinary Gaussian state. The choice \(u=-1,0,1\) gives
\begin{align}
&\Gamma_A^{1|2} = 2\Gamma_A(0) -\frac{1-\ii}{2}\Gamma_A(1) -\frac{1+\ii}{2}\Gamma_A(-1),\label{eq-bosonic-three-covariance-interpolation} \\ &\operatorname{Re}\Gamma_A^{1|2} = 2\Gamma_A(0) -\frac{\Gamma_A(1)+\Gamma_A(-1)}{2}, \qquad\operatorname{Im}\Gamma_A^{1|2} = \frac{\Gamma_A(1)-\Gamma_A(-1)}{2}.\label{eq-bosonic-three-covariance-real-imaginary}
\end{align}
When \(\Gamma_A(u)\) is independent of \(u\), one covariance matrix suffices. The full covariance remains polynomial of degree at most two in \(u\) after evolution generated by a quadratic Hamiltonian independent of \(u\). Three distinct real values of \(u\) therefore give the reduced covariance after evolution. One covariance matrix suffices only when this reduced covariance is independent of \(u\).

\subsection{Harmonic oscillators after a quench}
Let the ket and bra begin in oscillator ground states of possibly different frequencies and evolve with the same Hamiltonian. We show when their transition matrix equals the analytic continuation of the density-matrix formula for one ordinary Gaussian state to a complex time.

Let
\be
H_\omega\coloneqq\frac12\left(p^2+\omega^2q^2\right),\qquad[q,p]=\ii,\qquad\omega>0,
\ee
and let \(\ket{0_\omega}\) be its ground state. For \(\omega,\omega_1,\omega_2>0\), define
\be
\tau(t)\coloneqq\frac{e^{-\ii tH_\omega}\ket{0_{\omega_1}}\bra{0_{\omega_2}}e^{+\ii tH_\omega}}{\braket{0_{\omega_2}}{0_{\omega_1}}}.\label{eq-oscillator-transition-matrix}
\ee
Assume
\be
\omega>\max(\omega_1,\omega_2)\qquad\text{or}\qquad0<\omega<\min(\omega_1,\omega_2).\label{eq-oscillator-frequency-condition}
\ee
Equivalently, \(\lambda_1\lambda_2>0\). For \(i=1,2\),
\be
\lambda_i\coloneqq\frac{\omega-\omega_i}{\omega+\omega_i},\qquad\lambda_\Omega\coloneqq\sgn(\lambda_1)\sqrt{\lambda_1\lambda_2},\qquad\theta_\omega\coloneqq\frac{1}{4\omega}\log\!\left(\frac{\lambda_2}{\lambda_1}\right),\qquad\Omega\coloneqq\omega\frac{1-\lambda_\Omega}{1+\lambda_\Omega}.\label{eq-oscillator-frequency-and-time-shift}
\ee
The square root in \cref{eq-oscillator-frequency-and-time-shift} is the positive real root. Under \cref{eq-oscillator-frequency-condition}, \(\Omega>0\) and \(\theta_\omega\in\mathbb R\). Define
\be
\rho^{(\Omega)}(s)\coloneqq e^{-\ii sH_\omega}\proj{0_\Omega}e^{+\ii sH_\omega}.
\ee
Apart from \(\omega=\omega_1=\omega_2\), \cref{eq-oscillator-frequency-condition} is necessary and sufficient for \(\tau(t)=\rho^{(\Omega)}(t-\ii\theta_\omega)\) with \(\Omega>0\) and finite real \(\theta_\omega\). The matrix identity in \cref{eq-bosonic-transition-matrix-complex-parameters} holds for all \(\omega,\omega_1,\omega_2>0\). The proof of this frequency condition and of \cref{eq-oscillator-complex-time-identities} is given in \cref{app-oscillator}.

Let \(\Gamma^{1|2}(t)\) and \(\Gamma^{(\Omega)}(s)\) denote the covariance matrices of \(\tau(t)\) and \(\rho^{(\Omega)}(s)\), respectively. \Cref{app-oscillator} also includes the case of three equal frequencies and the cases in which one or both of \(\lambda_1,\lambda_2\) vanish, and gives the domain of holomorphy. Under \cref{eq-oscillator-frequency-condition}, it proves
\be
\frac{\ket{0_{\omega_1}}\bra{0_{\omega_2}}}{\braket{0_{\omega_2}}{0_{\omega_1}}}=e^{-\theta_\omega H_\omega}\proj{0_\Omega}e^{+\theta_\omega H_\omega},\qquad\tau(t)=\rho^{(\Omega)}(t-\ii\theta_\omega),\qquad\Gamma^{1|2}(t)=\Gamma^{(\Omega)}(t-\ii\theta_\omega).\label{eq-oscillator-complex-time-identities}
\ee
\paragraph{Several modes.} For a finite set \(\mathcal M\) of independent modes, let
\be
[q_\mu,p_\nu]=\ii\delta_{\mu\nu},\qquad H_\mu=\frac12\left(p_\mu^2+\omega_\mu^2q_\mu^2\right),\qquad\ket{0_j}=\bigotimes_{\mu\in\mathcal M}\ket{0_{\omega_{j,\mu}}},\qquad j=1,2.
\ee
Assume that each frequency triple \((\omega_\mu,\omega_{1,\mu},\omega_{2,\mu})\) satisfies \cref{eq-oscillator-frequency-condition}. Define \(\Omega_\mu,\theta_\mu\) by \cref{eq-oscillator-frequency-and-time-shift} and set
\begin{align}
&\tau(t)\coloneqq\frac{ e^{-\ii t\sum_\mu H_\mu} \ket{0_1}\bra{0_2} e^{+\ii t\sum_\mu H_\mu} }{ \braket{0_2}{0_1} }, & \qquad&\tau_A(t)\coloneqq\Tr_B\tau(t), \\ &\rho(\bm s)\coloneqq e^{-\ii\sum_\mu s_\mu H_\mu} \left( \bigotimes_{\mu\in\mathcal M} \proj{0_{\Omega_\mu}} \right) e^{+\ii\sum_\mu s_\mu H_\mu}, & &\rho_A(\bm s)\coloneqq\Tr_B\rho(\bm s).\label{eq-multimode-density-matrix}
\end{align}
Let \(\bm\theta\coloneqq(\theta_\mu)_{\mu\in\mathcal M}\) and \(\cD_{\bm\theta}\coloneqq\sum_{\mu\in\mathcal M}\theta_\mu\partial_{s_\mu}\). For real \(\bm s\), \(\rho_A(\bm s)\) is an ordinary reduced Gaussian density matrix. Define its entropy by \(S^{\mathrm E}(\bm s)\coloneqq S(\Gamma_A(\bm s))\). Applying \cref{eq-oscillator-complex-time-identities} to the independent modes gives the matrix and covariance identities below. The entropy equality also requires the conditions in \cref{app-bosonic-gaussian-kernel-covariance} and convergence of the entropy Taylor series at the displayed complex parameter values.
\be
\tau_A(t)=\left.\rho_A(\bm s)\right|_{s_\mu=t-\ii\theta_\mu},\quad\Gamma_A^{1|2}(t)=\left.\Gamma_A(\bm s)\right|_{s_\mu=t-\ii\theta_\mu},\quad S(\Gamma_A^{1|2}(t))=\left.e^{-\ii\cD_{\bm\theta}}S^{\mathrm E}(\bm s)\right|_{s_\mu=t,\ \mu\in\mathcal M}.\label{eq-multimode-complex-time-formulae}
\ee
The real and imaginary parts follow from \cref{eq-real-imaginary-taylor-series} with \(\cD_{\bm y}=-\cD_{\bm\theta}\). For the two Gaussian states at time \(t\), \cref{eq-entropy-average-and-half-difference,eq-excess-pseudo-entropy} give the average, half-difference, and excess using the real wavefunction parameters \((\mathsf X_i,\mathsf Y_i)\) and the derivatives in \cref{eq-bosonic-entropy-derivatives}, under the corresponding entropy and convergence conditions. If \(\theta_\mu=\theta\) for every mode, then
\be
\cD_{\bm\theta}=\theta\sum_\mu\partial_{s_\mu}=\theta\frac{\dd}{\dd s}\quad\text{on }s_\mu=s\text{ for all }\mu.
\ee
When the \(\theta_\mu\) differ, each variable \(s_\mu\) is evaluated at a different complex value.

\subsection{Free scalar lattice}\label{sec-scalar}
In real normal-mode coordinates, the free scalar lattice is a set of independent harmonic oscillators. Applying the preceding formulae to each mode expresses the covariance of the reduced transition matrix in terms of three ordinary Gaussian covariances for all positive masses. Writing the normal-mode transition matrices using separate complex times requires the mass condition stated below. The case of three equal masses is considered separately.

For the periodic \(N\)-site lattice regularisation used in \cite{Mollabashi:2021xsd}, let \(m>0\) be the mass and \(z\in\mathbb Z_{>0}\) the dynamical exponent. See also \cite{MohammadiMozaffar:2018vmk,Mozaffar:2021nex,Khoshdooni:2025ddk}. The normal-mode frequencies \cite{MohammadiMozaffar:2017nri,He:2017wla,Mollabashi:2021xsd}, with \(n=0,\ldots,N-1\), and the Hamiltonian written in real normal-mode coordinates are
\be
\omega_n(m,z)^2=m^{2z}+\left(2\sin\frac{\pi n}{N}\right)^{2z},\qquad H(m,z)=\sum_{\mu\in\mathcal M}\frac12\left[p_\mu^2+\omega_\mu(m,z)^2q_\mu^2\right].\label{eq-lattice-real-mode-Hamiltonian}
\ee
The real normal-mode decomposition is derived in \cref{app-scalar-real-modes}. For \(m,m_1,m_2>0\) and \(j=1,2\), define
\be
\begin{aligned} &\ket{0_j}\coloneqq\text{ground state of }H(m_j,z), & \qquad&\omega_{j,\mu}\coloneqq\omega_\mu(m_j,z), \\ &\omega_\mu\coloneqq\omega_\mu(m,z), & &\tau_A(t)\coloneqq\Tr_B\frac{ e^{-\ii tH(m,z)} \ket{0_1}\bra{0_2} e^{+\ii tH(m,z)} }{ \braket{0_2}{0_1} }. \end{aligned}
\ee
\paragraph{Interpolation from three covariance matrices.} Let \(\mathsf W_j\) be the positive matrix in the ground-state Gaussian wavefunction of \(H(m_j,z)\), and define
\be
\begin{aligned} &\mathsf W_{\mathrm c}\coloneqq\frac{\mathsf W_1+\mathsf W_2}{2},\qquad&&\mathsf W_\delta\coloneqq\frac{\mathsf W_2-\mathsf W_1}{2},\\ &\langle\bm q|\Psi(u)\rangle\propto\exp\!\left[-\frac12\bm q^{\mathsf T}(\mathsf W_{\mathrm c}+\ii u\mathsf W_\delta)\bm q\right],\qquad&&\rho(u,t)\coloneqq\frac{e^{-\ii tH(m,z)}\proj{\Psi(u)}e^{+\ii tH(m,z)}}{\braket{\Psi(u)}{\Psi(u)}}. \end{aligned}
\ee
These definitions agree with \cref{eq-bosonic-interpolation-family} after setting \(\mathsf X(u)=\mathsf W_{\mathrm c}\) and \(\mathsf Y(u)=u\mathsf W_\delta\). Since \(\mathsf W_{\mathrm c}>0\), every real \(u\) gives an ordinary Gaussian state. For complex \(u\), let \(\rho(u,t)\) and \(\Gamma_A(u,t)\) denote the analytic continuations of the kernel and its reduced covariance. At \(u=\ii\), the matrices in the exponents of the ket and bra Gaussian wavefunctions are \(\mathsf W_1\) and \(\mathsf W_2\), respectively. Applying \(e^{-\ii tH(m,z)}(\cdot)e^{+\ii tH(m,z)}\) and then \(\Tr_B\) to \cref{eq-bosonic-transition-matrix-complex-parameters}, the three-point interpolation in \cref{eq-bosonic-three-covariance-interpolation} gives, for all \(m_1,m_2>0\),
\be
\tau_A(t)=\Tr_B\rho(\ii,t),\qquad\Gamma_A^{1|2}(t)=2\Gamma_A(0,t)-\frac{1-\ii}{2}\Gamma_A(+1,t)-\frac{1+\ii}{2}\Gamma_A(-1,t).\label{eq-scalar-covariance-interpolation}
\ee
For real \(u\), \(S(\Gamma_A(u,t))\) is the ordinary Gaussian entropy. Under the conditions of \cref{app-bosonic-gaussian-kernel-covariance}, and when the Taylor series about \(u=0\) converges at \(u=\ii\), one obtains
\be
S(\Gamma_A^{1|2}(t))=\left.e^{+\ii\partial_u}S(\Gamma_A(u,t))\right|_{u=0}.
\ee
Using the real wavefunction parameters of the two states at time \(t\), \cref{eq-real-imaginary-taylor-series} gives the real and imaginary parts, and \cref{eq-entropy-average-and-half-difference,eq-excess-pseudo-entropy} give the average, half-difference, and excess, with the derivatives in \cref{eq-bosonic-entropy-derivatives} and their corresponding convergence conditions.

\paragraph{Mode-dependent complex times.} The case \(m=m_1=m_2\) is obtained by setting \(\Omega_\mu=\omega_\mu\) and \(\theta_\mu=0\) for every mode. Away from this case, applying \cref{eq-oscillator-complex-time-identities} to every normal mode is possible if and only if the following equivalent conditions hold.
\be
\begin{aligned} &m>\max(m_1,m_2)\qquad\text{or}\qquad0<m<\min(m_1,m_2)\\ &\quad\Longleftrightarrow\quad\left[\omega_\mu>\max(\omega_{1,\mu},\omega_{2,\mu})\qquad\text{or}\qquad0<\omega_\mu<\min(\omega_{1,\mu},\omega_{2,\mu})\right]\quad(\forall\mu). \end{aligned}\label{eq-scalar-frequency-condition}
\ee
The equivalence follows because the frequencies in \cref{eq-lattice-real-mode-Hamiltonian} are strictly increasing functions of the positive mass. Under \cref{eq-scalar-frequency-condition}, define \(\Omega_\mu\) and \(\theta_\mu\) for each mode by \cref{eq-oscillator-frequency-and-time-shift}. Together with the parameter values specified above when all three masses are equal, \cref{eq-multimode-complex-time-formulae} gives
\be
\tau_A(t)=\left.\rho_A(\bm s)\right|_{s_\mu=t-\ii\theta_\mu}.\label{eq-scalar-transition-matrix-complex-times}
\ee
Here \(\rho(\bm s)\) is defined in \cref{eq-multimode-density-matrix} with initial frequencies \(\Omega_\mu\). \Cref{eq-scalar-covariance-interpolation} holds for all \(m_1,m_2>0\), whereas \cref{eq-scalar-transition-matrix-complex-times} applies when \(m=m_1=m_2\) or when \cref{eq-scalar-frequency-condition} holds. The \(\Omega_\mu\) need not be of the form \(\omega_\mu(m,z)\) for one mass. Substituting these \(\Omega_\mu\) and \(\theta_\mu\) in \cref{eq-multimode-complex-time-formulae} gives the entropy of the free scalar lattice, whose real and imaginary parts follow from \cref{eq-real-imaginary-taylor-series}. \Cref{app-scalar-real-modes} proves \cref{eq-scalar-transition-matrix-complex-times} at finite \(N\) and compares the covariance matrices with those in \cite{Mollabashi:2021xsd}.

\subsection{Comparison with earlier quench calculations}
We compare numerical evaluations of the covariance formulae for the scalar lattice in \cref{eq-multimode-complex-time-formulae,eq-scalar-transition-matrix-complex-times} and the CFT entropy in \cref{eq-CFT-equal-time-entropy} with the quench examples in Figure 21 of \cite{Mollabashi:2021xsd}.
\begin{figure}[h!]
\centering\includegraphics[width=0.98\textwidth]{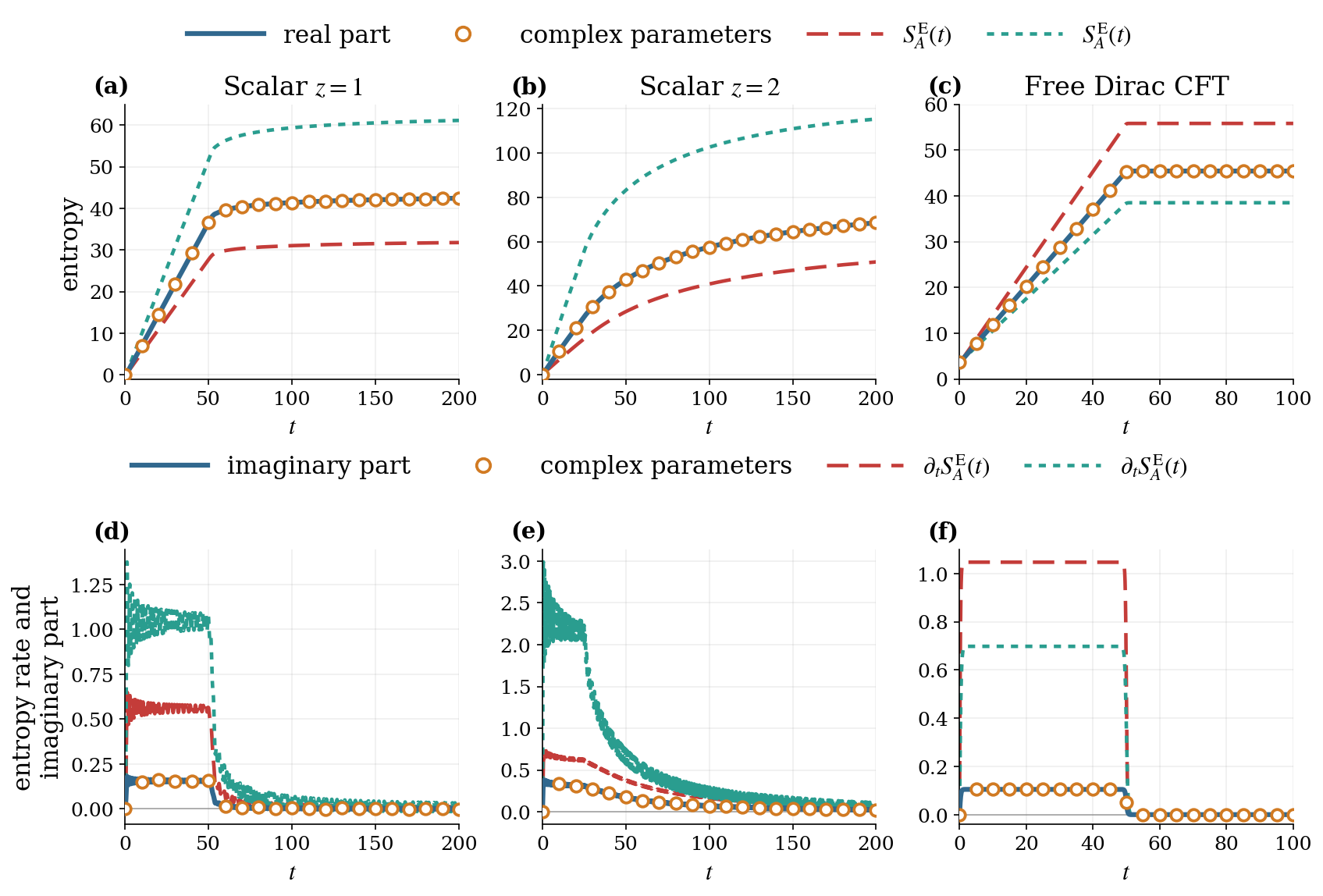} \caption{This figure reproduces the qualitative time dependence in Figure 21 of \cite{Mollabashi:2021xsd}. The three columns show the scalar models with \(z=1,2\) and the free Dirac CFT. The upper row shows the real part of pseudo entropy and the two entanglement entropies, and the lower row shows the imaginary part of pseudo entropy and the time derivatives of the two entanglement entropies. Hollow markers show the values for the scalar lattice obtained by evaluating the ordinary covariance of each mode at \(t-\ii\theta_n\), as in \cref{eq-scalar-transition-matrix-complex-times}, and the CFT values from \cref{eq-CFT-equal-time-entropy}. The time derivatives are shown for the qualitative comparison in \cite{Mollabashi:2021xsd}.} \label{fig-quench-comparison}
\end{figure}

For the scalar models, \(A\) is a contiguous block in a periodic chain of \(N=8192\) sites. The parameters are
\be
\begin{aligned} &\text{scalar}\qquad N_A=100,\quad m_1=2,\quad m_2=1,\quad m=10^{-5},\quad z=1,2, \\ &\text{CFT}\qquad\alpha_1=\frac12,\quad\alpha_2=\frac34,\quad\ell=100,\quad\eps=10^{-5}. \end{aligned}
\ee
The red and teal curves in \cref{fig-quench-comparison} correspond, respectively, to masses \(1\) and \(2\) for the scalar models and to \(\alpha_1\) and \(\alpha_2\) for the CFT.
\FloatBarrier

\section{Conclusions}
\paragraph{Summary.} An equality of matrices relates pseudo entropy to ordinary entanglement entropy. For two nonorthogonal states in a finite-dimensional Hilbert space, belonging to the same family with holomorphic coefficients, the reduced transition matrix is the ordinary reduced density-matrix formula evaluated at complex parameters fixed by the two states. This statement requires neither a matrix logarithm nor a Taylor series. Under the separate conditions for the entropy, its analytic continuation gives pseudo entropy. The values of ordinary entanglement entropy in the two chosen states do not suffice. Its parameter dependence is required. Near real parameters where all eigenvalues of the reduced density matrix are positive, the Taylor series in \cref{eq-transition-matrix-and-entropy-taylor-series} converges for sufficiently close states and gives their pseudo entropy.

The real and imaginary parts are given by the even and odd derivatives of the same entanglement-entropy function. Comparing the real part with the average entropy of the two states gives the excess pseudo entropy, which has no linear term in the differences between their parameter values. The corresponding R\'enyi relations require their own logarithm and convergence conditions. For states with polynomial coefficients, ordinary reduced density matrices and their normalisations at finitely many real parameter values give the reduced transition matrix and the traces of its fixed integer powers. For the family in \cref{eq-superposition-state-family}, the formulae in \cite{GuoZhang:2023sumrule,GuoJiangXu:2024} are related to this interpolation by a change of variables.

In boundary-state quenches, the matrix identity gives a complex change of time or of the parameter of Euclidean evolution. It reproduces the equal-time shift in \cite{Mollabashi:2021xsd} before calculating entropy. For the free Dirac CFT, the analytic continuation of the entropy gives an integral and a bound for its imaginary part. Under the assumptions of holomorphy in a half-plane and decay, the first moment gives a combination of the central charge and boundary one-point functions, and gives the central charge itself in the thermofield-double example. These entropy statements use the regulated limits and the logarithm branch fixed by continuation.

For fermionic Gaussian states with finitely many modes, polynomial coefficients allow interpolation using ordinary Gaussian states. For bosonic states, the kernel and covariance calculations replace the finite-dimensional argument, and the entropy is analytically continued using its covariance formula under the stated spectral conditions. The specified families require finitely many ordinary covariances together with known weights involving determinants. For the free scalar lattice, three covariances suffice for all positive masses. Writing the transition matrix instead as an ordinary density-matrix formula at complex times requires the additional frequency condition, or the case of equal initial and final frequencies in each mode. The times can differ between modes. Complex time is one realisation of the matrix relation, but the general matrix identity does not require it.

\paragraph{Future directions.} The main question is how to realise the continuation for a specified family of states in the gravitational description. At leading order, ordinary holographic entropy is given by areas of extremal surfaces \cite{RyuTakayanagi:2006,HubenyRangamaniTakayanagi:2007}, with related geometric descriptions for pseudo entropy and timelike entanglement \cite{Nakata:2020luh,Doi:2023Timelike,Heller:2024whi,Heller:2025kvp,Anastasiou:2025rvz,Anastasiou:2026bbf}. With the subsystem fixed, one should continue the bulk geometry and extremal surface in the state parameters while the bulk approximation remains valid. The thermal family in \cref{app-cft-tfd} provides a test using complex inverse temperature and the first-moment formula for the central charge. When several surfaces contribute, the dominant surface can change under continuation \cite{Heller:2025kvp}. The gravitational replica calculation should be compared with the analytic continuation of the entropy through these changes. When the area of the extremal surface gives the same holomorphic entropy, its real and imaginary parts obey the Cauchy-Riemann relations. The Kramers-Kronig relations also require the stated conditions on holomorphy in a half-plane and on decay. The entropy relations alone do not exclude saddles from the path integral.

Interactions and continuum limits provide a second direction. The finite-dimensional matrix identity applies with interactions, but removing regulators requires locally uniform convergence of the normalised reduced expressions and their entropies on a common complex domain, with normalisation and partial trace performed first. The interpolation formulae also need error bounds that retain the normalisations or weights involving determinants. These bounds must account for amplification of error by interpolation weights and small overlaps, and control the logarithm when computing entropy.

Within CFT, subtracting known contributions from operators of dimension below one may extend the first-moment relation to more boundary states. Integrals with higher powers of time likewise require subtracting terms that cause divergence and checking holomorphy and decay. Such integrals could relate further expansion coefficients to the imaginary part.

Although the imaginary part of pseudo entropy can be derived from entanglement entropy via analytic continuation, its direct quantum information-theoretic interpretation remains unclear. Given the interpretation of entanglement entropy in terms of Bell pairs distillable from pure states \cite{Horodecki:2009Entanglement}, the corresponding physical significance of this imaginary part remains a subject for future research.

\acknowledgments
We would like to thank Arindam Bhattacharjee, Abhirup Bhattacharya, Pawe\l{} Caputa, Avijit Das, Javier Moreno, and Karol {\.Z}yczkowski for their helpful comments and careful reviews of the draft. A.S. and P.S. are supported by OPUS grant no.~2022/\allowbreak47/B/ST2/03313 from the Narodowe Centrum Nauki (NCN), Poland. A.S. also acknowledges support from NCN SONATA BIS 9 grant no.~2019/34/E/ST2/00123, and thanks the Yukawa Institute for Theoretical Physics, Kyoto University, for its hospitality during a research visit supported by the travel microgrant BOB-IDUB-622-97-2026 from the University of Warsaw's IDUB Programme. T.T. is supported by Inamori Research Institute for Science and by JSPS Grant-in-Aid for Scientific Research (B) No.~25K01000.
\appendix
\crefalias{section}{appendix} \crefalias{subsection}{subappendix}

\section{Holomorphy, matrix functions, and integral formulae}
\subsection{Holomorphy and matrix functions}\label{app-holomorphy-and-taylor-series}
It is to be noted that holomorphy of the reduced matrix alone does not give holomorphy of its entropy. The contour and Taylor arguments below justify \cref{eq-entropy-taylor-series,eq-transition-matrix-and-entropy-taylor-series} under the stated logarithm and convergence conditions.
\paragraph{Holomorphy.} Let \(M\) be a square matrix and let \(f\) be holomorphic on a neighbourhood of its spectrum. Choose one or more positively oriented simple closed contours with disjoint interiors whose union \(\mathcal C\) encloses every eigenvalue exactly once, with each contour and its interior contained in the domain of \(f\) \cite{HighamLin:2013shortcourse}. Then
\be
f(M)=\frac{1}{2\pi\ii}\oint_{\mathcal C}f(\zeta)(\zeta\one-M)^{-1}\,\dd\zeta.\label{eq-matrix-function-contour}
\ee
For \(f=\Log\), every contour and its interior must lie in \(\mathbb C\setminus(-\infty,0]\). If \(M(\bm z)\) is holomorphic near \(\bm z_0\) and the same contours satisfy
\be
\operatorname{spec}M(\bm z)\subset\operatorname{int}\mathcal C\qquad(\bm z\ \text{near }\bm z_0),
\ee
then
\be
\begin{aligned} &M(\bm z)\ \text{holomorphic}\quad\Longrightarrow\quad(\zeta\one-M(\bm z))^{-1}\ \text{holomorphic}\quad\Longrightarrow\quad\Log M(\bm z)\ \text{holomorphic}\\ &\quad\Longrightarrow\quad-\Tr\!\left[M(\bm z)\Log M(\bm z)\right]\ \text{holomorphic}. \end{aligned}
\ee
\(S(\bm z)\) is holomorphic under the assumptions of \cref{sec-entropy-continuation}.

\paragraph{Taylor convergence at \(s=1\).} Set \(g(s)\coloneqq S(\bm x+\ii s\bm y)\). Under the more general assumptions on the segment and convergence stated before \cref{eq-entropy-taylor-series}, uniqueness of analytic continuation gives equality with the Taylor series for \(0\leq s<1\). If \(g(s)=\sum_{n\geq0}a_ns^n\) and \(A_n\coloneqq\sum_{k=0}^na_k\), then \(g(s)=(1-s)\sum_{n\geq0}A_ns^n\) for \(0\leq s<1\). Convergence of \(A_n\) implies \(\lim_{s\to1^-}g(s)=\sum_{n\geq0}a_n\). Continuity of \(g\) gives the same equality at \(s=1\).

\paragraph{Uniqueness.} Let \(\mathcal U\subset\mathbb C^d\) be connected, let \(\mathcal U_{\mathbb R}\subset\mathcal U\cap\mathbb R^d\) be nonempty and open in \(\mathbb R^d\), and let \(F,G\) be holomorphic on \(\mathcal U\). For any \(\bm x_0\in\mathcal U_{\mathbb R}\),
\be
\begin{aligned} &F|_{\mathcal U_{\mathbb R}}=G|_{\mathcal U_{\mathbb R}} \Longrightarrow(F-G)|_{\mathcal U_{\mathbb R}}=0\\ &\Longrightarrow\partial_{\bm x}^{\bm m}(F-G)(\bm x_0)=0 \Longrightarrow\partial_{\bm z}^{\bm m}(F-G)(\bm x_0)=0 \qquad(\bm m\in\mathbb N_0^d)\\ &\Longrightarrow F-G=0 \quad\text{in a complex neighbourhood of }\bm x_0 \Longrightarrow F-G=0 \quad\text{on }\mathcal U. \end{aligned}
\ee
The third implication uses holomorphy, and the last uses the identity theorem. For matrix-valued functions the argument applies entrywise. The holomorphic matrix-valued function in \cref{sec-entropy-continuation} is uniquely fixed by its values at real parameters.

For a finite-dimensional matrix \(K(\bm z)\), evaluation at \(\bm z_*\) commutes with linear maps between finite-dimensional matrix spaces, fixed integer powers and traces.
\be
\left.\mathcal L[K(\bm z)]\right|_{\bm z=\bm z_*}=\mathcal L[K(\bm z_*)],\qquad\left.K(\bm z)^n\right|_{\bm z=\bm z_*}=K(\bm z_*)^n,\qquad\left.\Tr[K(\bm z)^n]\right|_{\bm z=\bm z_*}=\Tr[K(\bm z_*)^n].
\ee
This includes partial trace, Fourier transforms whose coefficients do not depend on \(\bm z\), and taking the rows and columns belonging to a fixed subsystem. Evaluation also commutes with inverses and Schur complements wherever the required inverses exist. For a Gaussian integral, differentiation under the integral is valid on a neighbourhood if every derivative used is bounded there by one integrable function. For matrix functions, the same contours and logarithm branch must remain valid.
\paragraph{Hermiticity of the Taylor coefficients.} For the family in \cref{sec-two-states}, on a disc where the denominator in \cref{eq-holomorphic-matrix-family} is nonzero, define \(\tau_A(s)\coloneqq\rho_A(\bm t_{\mathrm c}+\ii s\bm\alpha_\delta,\bm\alpha_{\mathrm c}-\ii s\bm t_\delta)\). The definition \(\widetilde c_n(\bm z)=\overline{c_n(\overline{\bm z})}\) gives
\be
\rho_A(\bm t,\bm\alpha)^\dagger=\rho_A(\overline{\bm t},\overline{\bm\alpha})\quad\Longrightarrow\quad\tau_A(s)^\dagger=\tau_A(-\overline s),\qquad[\tau_A^{(k)}(0)]^\dagger=(-1)^k\tau_A^{(k)}(0),\qquad k\in\mathbb N_0.
\ee
The second identity follows by comparing Taylor coefficients. The even coefficients are Hermitian and the odd coefficients are anti-Hermitian, which proves the statement following \cref{eq-transition-matrix-and-entropy-taylor-series}.

\paragraph{Complex R\'enyi order.} For an initial pure state with reduced density matrix \(\rho_A\) whose eigenvalues are positive, consider evolution generated by \(-\log\rho_A\) on subsystem \(A\). The complex order \(w=1-\ii s\), with real \(s\), gives \(\Tr[\rho_A^{1-\ii s}]\) as the overlap of the evolved state with the initial state, as used in \cite{Caputa:2024ModularKrylov}. For the finite-dimensional family in \cref{sec-two-states}, the series of derivatives below uses the assumptions for the Taylor expansion in \cref{eq-matrix-function-taylor-series}. For fixed \(w\in\mathbb C\), let \(M=\rho_A(\bm t,\bm\alpha)\) or \(M=\tau_A\), assume that \(\Log M\) is defined and holomorphic throughout the complex parameter domain, and define
\be
M^w\coloneqq e^{w\Log M},\qquad Z_A(w,M)\coloneqq\Tr M^w,\qquad S_w(M)\coloneqq\frac{\Log Z_A(w,M)}{1-w},\qquad w\neq1.
\ee
Assume \(Z_A(w,M)\neq0\) and that one scalar logarithm branch with \(\Log1=0\) remains valid throughout the continuation. Then
\begin{align}
&Z_A(1,M)=1,\qquad&&\partial_w Z_A(w,M)=\Tr[M^w\Log M], \\ &\lim_{w\to1}S_w(M)=-\Tr[M\Log M],\qquad&&S_w(\tau_A)=\left.e^{\,\ii\cD_\perp}S_w(\rho_A(\bm t,\bm\alpha))\right|_{\bm t=\bm t_{\mathrm c},\,\bm\alpha=\bm\alpha_{\mathrm c}}.
\end{align}
The apparent singularity of \(S_w(M)\) at \(w=1\) is removable. The last equality follows from \cref{eq-matrix-function-taylor-series} because \(M\mapsto S_w(M)\) is holomorphic for fixed \(w\) under these assumptions. For \(n\in\mathbb Z_{\geq2}\), \(e^{n\Log M}=M^n\), so the definition for complex \(w\) agrees with the integer-order definition when the stronger matrix and scalar logarithm assumptions hold.

\paragraph{Trace-class operators.} Let
\be
\mathcal U\xrightarrow{\,T\,}\mathcal S_1(\mathcal H_A\otimes\mathcal H_B),\qquad\mathcal U\subset\mathbb C^d\ \text{open},\qquad T_A(\bm z)\coloneqq\Tr_BT(\bm z),\qquad\mathcal N(\bm z)\coloneqq\Tr T(\bm z).
\ee
Assume that \(T\) is holomorphic in trace norm, where \(\mathcal S_1\) is the trace class. Since partial trace and trace are bounded linear maps and \(\|XY\|_1\leq\|X\|_1\|Y\|_1\), on every connected component of \(\{\bm z\in\mathcal U\mid\mathcal N(\bm z)\neq0\}\),
\be
\rho_A(\bm z)\coloneqq\frac{T_A(\bm z)}{\mathcal N(\bm z)}\ \text{is holomorphic in trace norm},\qquad\bm z\longmapsto\Tr[\rho_A(\bm z)^n]\ \text{is holomorphic}.\label{eq-trace-class-extension}
\ee
Here \(n\in\mathbb Z_{>0}\). At a zero of \(\mathcal N\), \(T_A\) remains holomorphic, but \(\rho_A\) is undefined unless the quotient has a removable singularity. No entropy statement follows from \cref{eq-trace-class-extension} alone. An infinite-dimensional entropy additionally requires a common definition in terms of the spectrum with a convergent trace, or a regulated limit that converges locally uniformly on the same complex domain.

\subsection{Integral formulae}\label{app-integral-formulae}
We note that Taylor convergence alone is insufficient for integrals over all real times. Under the stated holomorphy and decay assumptions, contour integration gives \cref{eq-CFT-Kramers-Kronig} and the first moment of the imaginary part from the coefficient in \cref{eq-CFT-large-zeta-asymptotic}.
In this appendix, for a complex-valued function \(F\), write \(F^{\mathrm R}\coloneqq\operatorname{Re}F\) and \(F^{\mathrm I}\coloneqq\operatorname{Im}F\).

\paragraph{Integral formula on a horizontal strip.} Let \(a>0\), let \(F\) be holomorphic near the closed strip \(0\leq\operatorname{Im}z\leq a\), and assume \(F(x)\in\mathbb R\) for real \(x\). After subtracting a real constant if needed, assume that for every pair of nonnegative integers \(m,N\) there is a constant \(C_{m,N}\) such that \(|\partial_x^mF(x+\ii y)|\leq C_{m,N}(1+|x|)^{-N}\) uniformly for \(0\leq y\leq a\). These assumptions justify shifting the contour of the Fourier integral vertically within the strip and Fourier inversion. Writing \(F_a(x)\coloneqq F(x+\ii a)\) and \(\widehat h(k)=\int_{\mathbb R}e^{-\ii kx}h(x)\,\dd x\), Fourier transformation of the one-variable form of \cref{eq-real-imaginary-taylor-series} gives
\be
\widehat{F_a^{\mathrm R}}(k)=\cosh(ak)\widehat F(k),\qquad\widehat{F_a^{\mathrm I}}(k)=\ii\sinh(ak)\widehat F(k)=\ii\tanh(ak)\widehat{F_a^{\mathrm R}}(k).\label{eq-strip-fourier-relation}
\ee
With \(\operatorname{PV}\) denoting the Cauchy principal value, for \(k>0\) a lower-half-plane contour detouring below the pole at \(0\) gives
\be
\operatorname{PV}\!\int_{-\infty}^{\infty}e^{-\ii kx}\frac{1}{2a}\operatorname{csch}\!\left(\frac{\pi x}{2a}\right)\dd x=-2\ii\sum_{n\geq1}(-1)^ne^{-2akn}-\ii=-\ii\tanh(ak).\label{eq-csch-fourier-transform}
\ee
The case \(k<0\) follows by oddness. Combining \cref{eq-strip-fourier-relation,eq-csch-fourier-transform} gives
\be
F_a^{\mathrm I}(x)=\frac{1}{2a}\operatorname{PV}\!\int_{-\infty}^{\infty}\operatorname{csch}\!\left[\frac{\pi(s-x)}{2a}\right]F_a^{\mathrm R}(s)\,\dd s.
\ee
For fixed \(s-x\neq0\),
\be
\lim_{a\to\infty}\frac{1}{2a}\operatorname{csch}\!\left[\frac{\pi(s-x)}{2a}\right]=\frac{1}{\pi(s-x)}.
\ee
All functions below use a logarithm branch fixed by continuation from real parameter values. Real values on the relevant real axis give Schwarz reflection on the domain where holomorphy is assumed.
\paragraph{Integral of the imaginary part.} Let \(a\neq0\), and let \(F\) be holomorphic on an open set containing every \(x+\ii y\) with \(x\geq0\) and \(\min(0,a)\leq y\leq\max(0,a)\). Assume uniformly there that
\be
F(x)\in\mathbb R\qquad(x\geq0),\qquad F(z)=F_\infty+\mathcal O\!\left((1+\operatorname{Re}z)^{-1-\eta}\right),\qquad F_\infty\in\mathbb R,\quad\eta>0.
\ee
For \(a>0\), Cauchy's theorem on the rectangle with vertices \(0,\Lambda,\Lambda+\ii a,\ii a\) gives
\begin{align}
&0=\int_0^\Lambda F(x)\,\dd x+\ii\int_0^aF(\Lambda+\ii y)\,\dd y-\int_0^\Lambda F(x+\ii a)\,\dd x-\ii\int_0^aF(\ii y)\,\dd y,\\ &\Lambda\to\infty\quad\Longrightarrow\quad\int_0^\infty F^{\mathrm I}(t+\ii a)\,\dd t=\int_0^a\left[F_\infty-F^{\mathrm R}(\ii y)\right]\dd y.
\end{align}
For \(a<0\), use the oppositely oriented rectangle below the real axis.
\paragraph{Kramers-Kronig relations and the first moment of the imaginary part.} Let \(F(\zeta)\) be holomorphic for \(\operatorname{Re}\zeta>\alpha_*\), with \(\alpha_*\geq0\), and real for real \(\zeta>\alpha_*\). Assume that for some \(\eta>0\) and every fixed \(\alpha>\alpha_*\),
\be
F(\zeta)=\frac{C_2}{\zeta^2}+\mathcal O(|\zeta|^{-2-\eta}),\qquad\operatorname{Re}\zeta\geq\alpha,\label{eq-CFT-large-zeta-asymptotic}
\ee
uniformly in the indicated half-plane. Schwarz reflection gives
\be
C_2\in\mathbb R,\qquad F(\alpha-\ii t)=\overline{F(\alpha+\ii t)}.
\ee
For fixed \(\alpha>\alpha_*\), define \(G_\alpha(w)\coloneqq F(\alpha+\ii w)\). Since
\be
\operatorname{Im}w<0\quad\Longrightarrow\quad\operatorname{Re}(\alpha+\ii w)>\alpha>\alpha_*,
\ee
\(G_\alpha\) is holomorphic in the lower half-plane and \(G_\alpha(w)=\mathcal O(|w|^{-2})\). Cauchy's integral formula gives
\be
G_\alpha(t)=\frac{\ii}{\pi}\operatorname{PV}\!\int_{-\infty}^{\infty}\frac{G_\alpha(s)}{s-t}\,\dd s,\qquad t\in\mathbb R,
\ee
and therefore
\be
F^{\mathrm I}(\alpha+\ii t)=\frac1\pi\operatorname{PV}\!\int_{-\infty}^{\infty}\frac{F^{\mathrm R}(\alpha+\ii s)}{s-t}\,\dd s,\qquad F^{\mathrm R}(\alpha+\ii t)=-\frac1\pi\operatorname{PV}\!\int_{-\infty}^{\infty}\frac{F^{\mathrm I}(\alpha+\ii s)}{s-t}\,\dd s.
\ee
\Cref{eq-CFT-Kramers-Kronig} follows. Relations of the same form for thermal pseudo entropy were derived in \cite{Caputa:2024thermal}.

Applying Cauchy's integral formula to \(wF(\alpha+\ii w)\) and using Schwarz reflection gives
\begin{align}
&wF(\alpha+\ii w)=-\frac{C_2}{w}+\mathcal O\!\left(|w|^{-1-\min\{\eta,1\}}\right),\qquad\operatorname{PV}\!\int_{-\infty}^{\infty}wF(\alpha+\ii w)\,\dd w=-\ii\pi C_2,\\ &\operatorname{PV}\!\int_{-\infty}^{\infty}tF(\alpha+\ii t)\,\dd t=2\ii\int_0^\infty t\,F^{\mathrm I}(\alpha+\ii t)\,\dd t.
\end{align}
Therefore
\be
\int_0^\infty t\,F^{\mathrm I}(\alpha+\ii t)\,\dd t=-\frac\pi2C_2,\label{eq-CFT-first-moment}
\ee
which is independent of \(\alpha\). Expanding the second Kramers-Kronig relation for large \(t\), using Schwarz reflection and \cref{eq-CFT-large-zeta-asymptotic}, also gives \cref{eq-CFT-first-moment}. Under the CFT regulator and continuation conventions in \cref{sec-cft}, this identity may be written as an integral of the analytic continuation of pseudo entropy when the expression for the reduced transition matrix and the analytic continuation of its entropy exist throughout the integration range. Identifying it with the entropy defined using the principal matrix logarithm additionally requires agreement of the two definitions throughout that range.

\section{State families with polynomial coefficients and interpolation at real parameter values}
The unnormalised reduced matrix and its positive integer powers are polynomials, so interpolation gives the transition matrix and its integer moments from finitely many ordinary reduced density matrices and their normalisations. Evaluating the same polynomial by interpolation or coefficient extraction relates \cref{eq-three-density-matrix-formula} to earlier superposition formulae.
\subsection{Polynomial interpolation and coefficient formulae}\label{app-polynomial-interpolation}
Let
\be
\begin{aligned} &\ket{\phi(u)}\coloneqq\sum_I c_I(u)\ket{I},\qquad\widetilde c_I(u)\coloneqq\overline{c_I(\overline u)},\qquad\ket{\phi(\ii)}=\ket{\phi_1},\qquad\ket{\phi(-\ii)}=\ket{\phi_2},\\ &T_A(u)\coloneqq\Tr_B\!\left[\sum_{I,J}c_I(u)\widetilde c_J(u)\ket{I}\bra{J}\right],\quad\mathcal N(u)\coloneqq\Tr T_A(u), \quad\rho_A(u)\coloneqq\frac{T_A(u)}{\mathcal N(u)}\quad(u\in\mathbb R), \end{aligned}
\ee
Assume \(\deg c_I\leq r\), so \(\deg T_A,\deg\mathcal N\leq2r\). Also assume \(\mathcal N(u_k)\neq0\) at the real interpolation points and \(\braket{\phi_2}{\phi_1}\neq0\). Since \(T_A(\ii)=\braket{\phi_2}{\phi_1}\tau_A\) and \(\mathcal N(\ii)=\braket{\phi_2}{\phi_1}\), interpolation at \(2r+1\) distinct real points gives
\be
\tau_A=\frac{\displaystyle\sum_{k=0}^{2r}L_k^{(2r)}(\ii)\,\mathcal N(u_k)\rho_A(u_k)}{\displaystyle\sum_{k=0}^{2r}L_k^{(2r)}(\ii)\,\mathcal N(u_k)}.\label{eq-finite-polynomial-interpolation}
\ee
For \(n\geq2\), \(\deg T_A(u)^n\leq2nr\) and \(T_A(u)^n=\mathcal N(u)^n\rho_A(u)^n\) on the real axis. Lagrange interpolation at \(2nr+1\) distinct real points gives
\be
\Tr\!\left[\left(\tau_A\right)^n\right]=\frac{\displaystyle\sum_{k=0}^{2nr}L_k^{(2nr)}(\ii)\,\mathcal N(u_k)^n\Tr[\rho_A(u_k)^n]}{\displaystyle\left[\sum_{k=0}^{2nr}L_k^{(2nr)}(\ii)\,\mathcal N(u_k)\right]^n}.\label{eq-finite-polynomial-moment}
\ee
Neither \cref{eq-finite-polynomial-interpolation,eq-finite-polynomial-moment} contains a logarithm or branch choice.

For a polynomial \(P\), Cauchy's formula and evaluations at roots of unity also give \(P(\ii)\). For the linear superposition family, the change of variables is given in \cref{eq-superposition-parameter-change,eq-root-of-unity-operator-sum}.

\subsection{Formula from three density matrices}\label{app-superposition-sum-rule}
\paragraph{Values at \(u=-1,0,1\).} For the family in \cref{eq-superposition-state-family}, define \(T_A(u)\coloneqq\mathcal N(u)\rho_A(u)\). We write \([u^k]P(u)\) for the coefficient of \(u^k\) in a polynomial \(P(u)\). The three Lagrange weights below are labelled by their interpolation points \(-1,0,1\). Then
\begin{align}
&T_A(u)=T_{A,0}+uT_{A,1}+u^2T_{A,2}, \\ &T_{A,0}\coloneqq\Tr_B\proj{\varphi_{\mathrm c}},\qquad T_{A,2}\coloneqq\Tr_B\proj{\varphi_\delta}, \qquad T_{A,1}\coloneqq\ii\Tr_B\!\left(\ket{\varphi_\delta}\bra{\varphi_{\mathrm c}}-\ket{\varphi_{\mathrm c}}\bra{\varphi_\delta}\right), \\ &L_{-1}^{(2)}(\ii)=-\frac{1+\ii}{2},\qquad L_0^{(2)}(\ii)=2,\qquad L_{+1}^{(2)}(\ii)=-\frac{1-\ii}{2}.\label{eq-three-point-lagrange-weights}
\end{align}
Since \(T_A(\ii)=\braket{\psi_2}{\psi_1}\tau_A\) and \(\Tr T_A(\ii)=\braket{\psi_2}{\psi_1}\), inserting \cref{eq-three-point-lagrange-weights} into \cref{eq-finite-polynomial-interpolation} with \(r=1\) proves \cref{eq-three-density-matrix-formula}. The points \(-1,0,1\) are one choice for interpolating a quadratic polynomial. The three roots of unity used in \cite{GuoZhang:2023sumrule} give the points obtained below. If the two states are not nonzero scalar multiples of one another, \(T_{A,2}\neq0\), and for every integer \(n\geq1\),
\begin{align}
&[u^{2n}]T_A(u)^n=T_{A,2}^n\neq0, \qquad[u^{2n}]\Tr[T_A(u)^n]=\Tr(T_{A,2}^n)>0, \\ &\deg T_A(u)^n=\deg\Tr[T_A(u)^n]=2n.\label{eq-linear-superposition-moment-degree}
\end{align}
The values of the normalisation and of \(\Tr[T_A(u)^n]\) at \(2n+1\) real interpolation points give \(\Tr[(\tau_A)^n]\).
\paragraph{Comparison with earlier formulae.} The family in \cref{eq-superposition-state-family} can be written as
\begin{align}
&\ket{\varphi(u)}=\frac{1+\ii u}{2}\left(\ket{\psi_2}+\lambda(u)\ket{\psi_1}\right),\qquad\lambda(u)\coloneqq\frac{1-\ii u}{1+\ii u},\label{eq-superposition-parameter-change}\\ &\text{For }u\in\mathbb R,\qquad|\lambda(u)|=1,\qquad\lambda(u)=e^{\ii\vartheta}\Longleftrightarrow u=-\tan(\vartheta/2), \\ &m\coloneqq2n+1,\qquad\lambda_k\coloneqq e^{2\pi\ii k/m},\qquad k=0,\ldots,2n, \\ &u_k\coloneqq\ii\frac{\lambda_k-1}{\lambda_k+1}=-\tan\!\left(\frac{\pi k}{m}\right)\in\mathbb R,\qquad\left(\frac{1+u_k^2}{4}\right)^nL_k^{(2n)}(\ii)=\frac{\lambda_k^{-n}}{2n+1}.\label{eq-root-of-unity-lagrange-weights}
\end{align}
Let
\be
\ket{\psi(\lambda)}=\mathcal N(\lambda)\left(\ket{\psi_2}+\lambda\ket{\psi_1}\right),\qquad\mathcal N(\lambda)>0,\qquad\braket{\psi(\lambda)}{\psi(\lambda)}=1,
\ee
and let \(\rho_A(\lambda)\) be its reduced density matrix. Using \cref{eq-root-of-unity-lagrange-weights} in the Lagrange formula gives
\be
(\tau_A)^n=\frac{1}{(2n+1)\braket{\psi_2}{\psi_1}^{\,n}}\sum_{k=0}^{2n}\lambda_k^{-n}\mathcal N(\lambda_k)^{-2n}\rho_A(\lambda_k)^n.\label{eq-root-of-unity-operator-sum}
\ee
With the change of variables and weights in \cref{eq-root-of-unity-lagrange-weights}, the formula in \cite{GuoZhang:2023sumrule} evaluates the polynomial \(T_A(u)^n\), whose degree is given in \cref{eq-linear-superposition-moment-degree}, at \(u=\ii\), using the real points \(u_k\) corresponding to the roots of unity \(\lambda_k\). Taking the trace gives the relation for pseudo-R\'enyi entropies. Coefficient extraction gives the contour formula in \cite{GuoJiangXu:2024}. The Lagrange formula allows any distinct real interpolation points. These identities hold for each fixed integer \(n\) and do not give a \(w\to1\) continuation. If the norm of a superposition vanishes, its normalised density matrix is undefined. In that case use the unnormalised polynomial identity before division.

\section{Conformal field theory calculations}
\subsection{Free Dirac boundary-state formulae} \label{app-cft-free-dirac}
To justify \cref{eq-CFT-equal-time-entropy}, we prove holomorphy and Taylor convergence for the free Dirac entropy separately from the equality of integer moments. This entropy also gives the integral bound and first moment in \cref{eq-CFT-imaginary-part-integral-bound,eq-free-dirac-first-moment}.
\paragraph{Conformal map and moments for integer \(n\).} With \(\alpha_{\mathrm c}\) and \(\alpha_\delta\) from \cref{eq-parameter-centre-and-half-difference}, the conformal map used in \cite{Mollabashi:2021xsd} for the equal-time transition matrix in \cref{eq-CFT-equal-time-complex-parameter} satisfies
\be
\begin{aligned} &f_{\mathrm c}(z)\coloneqq\exp\!\left(\frac{\pi z}{2\alpha_{\mathrm c}}\right),\qquad f_{12}(z)=f_{\mathrm c}(z-\ii\alpha_\delta),\qquad f'_{12}(z)=f'_{\mathrm c}(z-\ii\alpha_\delta),\\ &\bar f_{12}(\bar z)=\bar f_{\mathrm c}(\bar z+\ii\alpha_\delta),\qquad\bar f'_{12}(\bar z)=\bar f'_{\mathrm c}(\bar z+\ii\alpha_\delta),\qquad t'=t+\ii\alpha_\delta. \end{aligned}
\ee
For every fixed integer \(n\geq2\), the coordinates of the twist operators and the Jacobian factors agree with those of the ordinary state evaluated at \(t'\). Thus
\be
\Tr\!\left[(\tau_A(t))^n\right]=\left.\Tr\!\left[\rho_A(t',\alpha_{\mathrm c})^n\right]\right|_{t'=t+\ii\alpha_\delta}.\label{eq-CFT-replica-moments}
\ee
For every fixed integer \(n\), taking the \(n\)-th power and trace of the matrix equality in \cref{eq-CFT-equal-time-complex-parameter} gives \cref{eq-CFT-replica-moments}, which agrees with the calculation using the conformal map. These moments of integer order alone do not give an analytic continuation in \(w\) or the logarithm branch needed for the entropy. The limit \(\eps\to0\) is taken for the vacuum-subtracted entropy, or for traces of powers of the reduced matrix divided by the same traces for the vacuum at the same regulator, with locally uniform convergence on the same complex parameter domain. When a derivative with respect to \(w\) is used, a holomorphic continuation near \(n=1\) and locally uniform convergence jointly in that neighbourhood and the complex parameters are required. Convergence at integer orders alone is insufficient.
\paragraph{Free Dirac entropy and its derivatives.} Since \(\alpha_1,\alpha_2>0\), \(|\alpha_\delta|<\alpha_{\mathrm c}\). The singularities in \(t'\) have imaginary parts \((2k+1)\alpha_{\mathrm c}\), \(k\in\mathbb Z\). Hence the logarithm continued from real \(t'\) is holomorphic on a disc containing \(t'=t+\ii\alpha_\delta\). Substituting \(t'=t+\ii\alpha_\delta\) in \cref{eq-CFT-entanglement-entropy} gives the free-Dirac pseudo entropy in \cite{Mollabashi:2021xsd}. Differentiating \cref{eq-CFT-equal-time-entropy} with respect to \(\alpha_\delta\) at fixed \(\alpha_{\mathrm c}\) gives
\be
\left.\frac{\partial^m}{\partial\alpha_\delta^m}S(\tau_A(t))\right|_{\alpha_\delta=0}=\ii^m\partial_t^mS^{\mathrm E}(t,\alpha_{\mathrm c}).
\ee
Separating even and odd \(m\) reproduces the expansions in \cite{Mollabashi:2021xsd}. Holomorphy on the domain above proves convergence of these Taylor series for \(\alpha_1,\alpha_2>0\).

For fixed real \(t\), the argument of the logarithm in \cref{eq-CFT-entanglement-entropy} has no zeros or poles for \(\operatorname{Re}\alpha>0\). Its logarithm continued from positive real \(\alpha\) is therefore holomorphic there. Since \(|\alpha_\delta|<\alpha_{\mathrm c}\), its Taylor series about \(\alpha=\alpha_{\mathrm c}\) converges at \(\alpha_{\mathrm c}\pm\alpha_\delta\).

At fixed integer \(n\), the temporal derivatives can also be written as the nested commutators used in \cite{GuoHeZhang:2024TimelikeSpacelike,XuGuo:2025Imaginary}. \Cref{eq-CFT-equal-time-entropy} gives the entropy without introducing those commutators.
\paragraph{Integral and first-moment formulae.} Taking \(F(z)=S(z,\alpha_{\mathrm c})\) in \cref{app-integral-formulae} and using \cref{eq-CFT-entanglement-entropy} proves \cref{eq-CFT-imaginary-part-integral}. Set \(L\coloneqq\pi\ell/(4\alpha_{\mathrm c})\) and \(J(L)\coloneqq\int_0^{\pi/2}\log[1+\sinh^2L/\cos^2u]\,\dd u\). With \(x=\tan u\),
\be
J'(L)=2\sinh L\cosh L\int_0^\infty\frac{\dd x}{\cosh^2L+\sinh^2L\,x^2}=\pi,\qquad J(0)=0\quad\Longrightarrow\quad J(L)=\pi L.
\ee
\Cref{eq-CFT-imaginary-part-integral-bound} follows. Expanding \cref{eq-CFT-imaginary-part-integral} at \(\alpha_\delta=0\) and using \cref{eq-CFT-entanglement-entropy} gives
\be
\int_0^\infty\SI(\tau_A(t))\,\dd t=\alpha_\delta\left[S^{\mathrm E}(\infty,\alpha_{\mathrm c})-S^{\mathrm E}(0,\alpha_{\mathrm c})\right]+\mathcal O(\alpha_\delta^3)=\frac{\alpha_\delta}{3}\log\cosh\!\left(\frac{\pi\ell}{4\alpha_{\mathrm c}}\right)+\mathcal O(\alpha_\delta^3).
\ee
For the free Dirac function in \cref{eq-CFT-vacuum-subtracted-entropy}, expanding \(\tanh x/x=1-x^2/3+\mathcal O(x^4)\) at \(x=0\) gives
\be
F(\zeta)=-\frac{\pi^2\ell^2}{144\zeta^2}+\mathcal O(\zeta^{-4}),\qquad\alpha_*=0,\qquad C_2=-\frac{\pi^2\ell^2}{144}.
\ee
The argument of the logarithm has no zeros or poles for \(\operatorname{Re}\zeta>0\), so its logarithm continued from positive real \(\zeta\) is holomorphic there. \Cref{eq-CFT-first-moment} gives \cref{eq-free-dirac-first-moment}.

\subsection{Short-interval expansion and the first moment}\label{app-cft-short-interval-first-moment}
To derive \cref{eq-CFT-boundary-first-moment}, we need the coefficient of \(\zeta^{-2}\) in the vacuum-subtracted entropy. The short-interval expansion gives the contributions from the central charge and from dimension-one operators.
For a unitary two-dimensional CFT on the infinite line with conformal boundary state \(\ket{\mathcal B}\), the path integral for \(e^{-\zeta H}\ket{\mathcal B}\), \(\zeta>0\), is defined on a strip of width \(2\zeta\) \cite{Guo:2018Boundary}. For \(A=[-\ell/2,\ell/2]\), define
\be
F_{\mathcal B}(\zeta,\ell)\coloneqq S_{A,\mathcal B}^{\mathrm E}(0,\zeta)-S_{A,\mathrm{vac}}.
\ee
Here \(S_{A,\mathrm{vac}}\) is the vacuum entanglement entropy of the same interval in the same additive normalisation. We use \(F_{\mathcal B}\) for the analytic continuation of this vacuum-subtracted expression to complex \(\zeta\).

For non-identity Hermitian spinless bulk primaries \(\mathcal O_a\) of dimensions \(\Delta_a\), use the normalisation and one-point functions on the strip given in \cite{Guo:2018Boundary} and the energy density of \cite{LencsesVitiTakacs:2019}.
\be
\langle\mathcal O_a(z,\bar z)\mathcal O_b(0)\rangle=\frac{\delta_{ab}}{|z|^{2\Delta_a}},\qquad\langle\mathcal O_a\rangle_{\mathcal B,\zeta}=b_a^{(\mathcal B)}\left(\frac{\pi}{4\zeta}\right)^{\Delta_a},\qquad\langle T_{00}\rangle_{\mathcal B,\zeta}=\frac{\pi c}{96\zeta^2},
\ee
where \(b_a^{(\mathcal B)}\in\mathbb R\) for real \(\zeta>0\). Let \(\rho_A^{(0)}\) be the vacuum reduced density matrix, define \(D(\rho\Vert\sigma)\coloneqq\Tr[\rho(\log\rho-\log\sigma)]\), and let \(\Delta\langle T_{00}\rangle\) and \(\Delta S_A\) denote the differences from their vacuum values. The vacuum modular Hamiltonian \cite{CasiniHuertaMyers:2011} and the relative-entropy expansion for scalar primaries \cite{SarosiUgajin:2017Relative} give
\be
\begin{aligned} &K_A^{(0)} = 2\pi\int_{-\ell/2}^{\ell/2} \frac{(\ell/2)^2-x^2}{\ell} T_{00}(x)\,\dd x +\text{constant}, & &\Delta\langle K_A^{(0)}\rangle= \frac{\pi\ell^2}{3}\, \Delta\langle T_{00}\rangle, \\ &D(\rho_A\Vert\rho_A^{(0)}) = \Delta\langle K_A^{(0)}\rangle-\Delta S_A = \sum_a \kappa_{\Delta_a} \left(\frac{\ell}{2}\right)^{2\Delta_a} \langle\mathcal O_a\rangle^2 +\cdots, & &\kappa_\Delta= \frac{\sqrt{\pi}\,\Gamma(\Delta+1)} {4\,\Gamma(\Delta+\tfrac32)}. \end{aligned}
\ee
Hence
\be
F_{\mathcal B}(\zeta,\ell)=\frac{c\pi^2\ell^2}{288\zeta^2}-\sum_a\kappa_{\Delta_a}\left(\frac{\pi\ell}{8\zeta}\right)^{2\Delta_a}\bigl(b_a^{(\mathcal B)}\bigr)^2+\cdots.\label{eq-CFT-short-interval-expansion}
\ee
Assume
\be
b_a^{(\mathcal B)}\neq0\ \Longrightarrow\ \Delta_a\geq1,\qquad\Sigma_{1,\mathcal B}\coloneqq\sum_{\Delta_a=1}\bigl(b_a^{(\mathcal B)}\bigr)^2\geq0.
\ee
If the omitted terms obey the uniform remainder bound in \cref{eq-CFT-large-zeta-asymptotic}, \cref{eq-CFT-short-interval-expansion} with \(\kappa_1=1/3\) gives
\be
F_{\mathcal B}(\zeta,\ell)=\frac{\pi^2\ell^2}{288\zeta^2}\left(c-\frac32\Sigma_{1,\mathcal B}\right)+\mathcal O(|\zeta|^{-2-\eta}).
\ee
If \(F_{\mathcal B}\) is holomorphic in a right half-plane containing \(\operatorname{Re}\zeta=\alpha\), applying \cref{eq-CFT-first-moment} to the preceding asymptotic expansion gives
\be
-\frac{576}{\pi^3\ell^2}\int_0^\infty t\,F_{\mathcal B}^{\mathrm I}(\alpha+\ii t,\ell)\,\dd t=c-\frac32\Sigma_{1,\mathcal B}.
\ee
Under the branch and transition-matrix conditions of \cref{app-integral-formulae}, analytic continuation gives
\be
F_{\mathcal B}(\alpha+\ii t,\ell)=S(\tau_{A,\mathcal B}(t,\alpha))-S_{A,\mathrm{vac}},\qquad F_{\mathcal B}^{\mathrm I}(\alpha+\ii t,\ell)=\SI(\tau_{A,\mathcal B}(t,\alpha)).
\ee
Substituting the preceding identities into \cref{eq-CFT-first-moment} proves \cref{eq-CFT-boundary-first-moment}. Setting \(\Sigma_{1,\mathcal B}=0\) proves \cref{eq-CFT-central-charge-first-moment}.

Contributions from primaries with \(\Delta_a>1\) decay faster than \(\zeta^{-2}\), while any nonzero \(\Delta_a<1\) one-point function would violate the asymptotic condition in \cref{eq-CFT-large-zeta-asymptotic}.

\subsection{Coefficients in one-point functions of dimension-one operators in the free Dirac CFT}\label{app-free-dirac-dimension-one-coefficients}
We evaluate the coefficients entering \(\Sigma_{1,\mathcal B}\) for the standard linear boundary conditions of the free Dirac fermion, using the local Neveu-Schwarz operator sector that does not change the spin structure, so Ramond spin fields are not included in the sum. This gives \(\Sigma_{1,\mathcal B}\) independently of the entropy calculation, so \cref{eq-CFT-boundary-first-moment} can be compared with \cref{eq-free-dirac-first-moment}.

Write the free Dirac fermion as two real Majorana fermions \(\chi^p\), \(p=1,2\), normalised by
\be
\chi_{\mathrm L}^{p}(z)\chi_{\mathrm L}^{r}(0)\sim\frac{\delta^{pr}}{z},\qquad\chi_{\mathrm R}^{q}(\bar z)\chi_{\mathrm R}^{s}(0)\sim\frac{\delta^{qs}}{\bar z}.
\ee
A real linear conformal boundary condition can be written
\be
\chi_{\mathrm L}^{p}(x)=\mathsf G_{pq}\chi_{\mathrm R}^{q}(x),\qquad\mathsf G\mathsf G^{\mathsf T}=\one_2,\qquad\mathsf G\in O(2).
\ee
For a single Dirac fermion, the two families of boundary conditions in \cite{SmithTong:2020BoundaryStates} are contained in this form.
\be
\det\mathsf G=+1\quad\Longleftrightarrow\quad\psi_{\mathrm L}=e^{\ii\theta}\psi_{\mathrm R},\qquad\det\mathsf G=-1\quad\Longleftrightarrow\quad\psi_{\mathrm L}=e^{\ii\theta}\psi_{\mathrm R}^{\dagger}.
\ee
These are the standard Neumann and Dirichlet families in the bosonised description, including their boundary phases. Denote the boundary condition specified by \(\mathsf G\) by \(\mathcal B_{\mathsf G}\).

The Hermitian spinless dimension-one primaries in this operator sector are
\be
\mathcal O_{pq}(z,\bar z)\coloneqq\ii\chi_{\mathrm L}^{p}(z)\chi_{\mathrm R}^{q}(\bar z),\qquad\left\langle\mathcal O_{pq}(z,\bar z)\mathcal O_{rs}(0)\right\rangle=\frac{\delta_{pr}\delta_{qs}}{|z|^2},\qquad p,q=1,2.
\ee
They are complete in this sector because a spinless operator of total dimension one has \((h,\bar h)=(1/2,1/2)\), and each space of chiral operators of weight \(1/2\) is spanned by the two Majorana fermions.

The doubling trick and the conformal map \(u=\ii\exp[\pi w/(2\zeta)]\) from the strip give
\begin{align}
&\left\langle\mathcal O_{pq}(z,\bar z)\right\rangle_{\mathcal B_{\mathsf G}} =\frac{\mathsf G_{pq}}{2\operatorname{Im}z},\qquad\left\langle\mathcal O_{pq}\right\rangle_{\mathcal B_{\mathsf G},\zeta} =\mathsf G_{pq}\frac{\pi}{4\zeta},\\ &b_{pq}^{(\mathcal B_{\mathsf G})}=\mathsf G_{pq},\qquad\Sigma_{1,\mathcal B_{\mathsf G}} =\sum_{p,q=1}^{2}\left(b_{pq}^{(\mathcal B_{\mathsf G})}\right)^2 =\Tr\!\left(\mathsf G\mathsf G^{\mathsf T}\right)=2.
\end{align}
Every boundary condition in these two families gives \(\Sigma_{1,\mathcal B_{\mathsf G}}=2\), which agrees with \cref{eq-free-dirac-first-moment,eq-CFT-boundary-first-moment}. Including local fields that change the spin structure, or using other \(c=1\) boundary conditions, requires a separate evaluation of \(\Sigma_{1,\mathcal B}\).

\subsection{First-moment identity for the thermofield double} \label{app-cft-tfd}
The vacuum-subtracted interval entropy in the infinite-line limit satisfies the holomorphy and decay conditions for \cref{eq-CFT-first-moment}. Its \(\zeta_\beta^{-2}\) coefficient gives the central charge in \cref{eq-TFD-first-moment-central-charge}. For two thermofield-double states, use the parameters in \cref{eq-TFD-parameters}. Their overlap and the transition matrix reduced to the right copy are
\be
\left\langle\operatorname{TFD}(\beta_2,t_2)\middle|\operatorname{TFD}(\beta_1,t_1)\right\rangle=\frac{Z(\zeta_\beta)}{\sqrt{Z(\beta_1)Z(\beta_2)}},\qquad\tau_{\mathrm R}=\frac{e^{-\zeta_\beta H_{\mathrm R}}}{Z(\zeta_\beta)}.
\ee
For \cref{eq-CFT-first-moment}, assume \(Z(\beta_{\mathrm c}+\ii s)\neq0\) for \(s\geq0\). The TFD transition matrix is the analytic continuation of the ordinary thermal density matrix from real inverse temperature to \(\zeta_\beta=\beta_{\mathrm c}+\ii s\), as studied in \cite{GotoNozakiTamaoka:2021,Caputa:2024thermal}.

For the interval calculation, take the infinite-line limit before the large-\(s\) limit and the integral below. In this limit, for an interval \(A\) of length \(\ell\) in the right copy, continuation from real \(\zeta_\beta>0\), with the logarithm branch fixed by that continuation, gives \cite{CalabreseCardy:2004,GotoNozakiTamaoka:2021}
\be
S(\zeta_\beta)=\frac{c}{3}\log\!\left[\frac{\zeta_\beta}{\pi\eps}\sinh\!\left(\frac{\pi\ell}{\zeta_\beta}\right)\right],\quad F_\beta(\zeta_\beta)\coloneqq S(\zeta_\beta)-\frac{c}{3}\log\!\left(\frac{\ell}{\eps}\right)=\frac{c}{3}\log\!\left[\frac{\zeta_\beta}{\pi\ell}\sinh\!\left(\frac{\pi\ell}{\zeta_\beta}\right)\right].
\ee
The argument of the logarithm in \(F_\beta(\zeta_\beta)\) has finite zeros only at \(\zeta_\beta=-\ii\ell/k\), \(k\in\mathbb Z\setminus\{0\}\). It is nonzero for \(\operatorname{Re}\zeta_\beta>0\), and the logarithm continued from positive real \(\zeta_\beta\) is holomorphic there.
\be
F_\beta(\zeta_\beta)=\frac{c\pi^2\ell^2}{18\zeta_\beta^2}+\mathcal O(\zeta_\beta^{-4}),\qquad C_2=\frac{c\pi^2\ell^2}{18}.
\ee
The vacuum subtraction is real, so \(F_\beta^{\mathrm I}(\zeta_\beta)=S^{\mathrm I}(\zeta_\beta)\), and \cref{eq-CFT-first-moment} gives \cref{eq-TFD-first-moment-central-charge}.

\section{Calculations for Gaussian states and quenches}
\subsection{Polynomial degree and operator matrix elements for fermionic Gaussian states}\label{app-fermion-interpolation}
The degree bounds proved here give sufficient numbers of interpolation points for the family in \cref{eq-fermion-polynomial-family}. Ordinary Gaussian states and their normalisations at these points give the reduced transition matrix, its moments of integer order, and operator matrix elements.
\paragraph{Polynomial degree.} For the family \cref{eq-fermion-polynomial-family}, define
\be
\mathcal P_{\mathrm c}\coloneqq\frac12\sum_{i,j=1}^{N}\beta_i^\dagger(\mathsf Z_{\mathrm c})_{ij}\beta_j^\dagger,\qquad\mathcal P_\delta\coloneqq\frac12\sum_{i,j=1}^{N}\beta_i^\dagger(\mathsf Z_\delta)_{ij}\beta_j^\dagger.
\ee
The two polynomials contain only products of two creation operators and commute. If \(\operatorname{rank}\mathsf Z_\delta=2r\), a change of one-particle basis puts \(\mathsf Z_\delta\) in canonical antisymmetric form, so
\be
\mathcal P_\delta=\sum_{a=1}^{r}\lambda_ad_{2a-1}^\dagger d_{2a}^\dagger,\qquad\lambda_a\neq0.
\ee
Hence
\be
\mathcal P_\delta^{\,r}\ket{\Phi_0}=r!\left(\prod_{a=1}^{r}\lambda_a\right)d_1^\dagger\cdots d_{2r}^\dagger\ket{\Phi_0}\neq0,\qquad\mathcal P_\delta^{\,r+1}=0.
\ee
Define
\be
\ket{v_k}\coloneqq\frac{\ii^k}{k!}e^{\mathcal P_{\mathrm c}}\mathcal P_\delta^k\ket{\Phi_0},\qquad\ket{\widetilde\Psi(\mathsf Z(u))}=e^{\mathcal P_{\mathrm c}}\sum_{k=0}^{r}\frac{(\ii u\mathcal P_\delta)^k}{k!}\ket{\Phi_0}=\sum_{k=0}^{r}u^k\ket{v_k}.
\ee
Since \(e^{\mathcal P_{\mathrm c}}\) is invertible, \(\ket{v_r}\neq0\). Define \(T_A(u)\coloneqq\mathcal N(u)\rho_A(u)\) and \(T_{A,2r}\coloneqq\Tr_B\proj{v_r}\). Then \(T_{A,2r}\geq0\) and \(\Tr T_{A,2r}=\braket{v_r}{v_r}>0\), so
\be
[u^{2r}]T_A(u)=T_{A,2r}\neq0,\qquad[u^{2r}]\mathcal N(u)=\braket{v_r}{v_r}>0.
\ee
Thus \(\deg T_A=\deg\mathcal N=2r\) and
\be
[u^{2nr}]\Tr[T_A(u)^n]=\Tr(T_{A,2r}^n)>0,\qquad\deg\Tr[T_A(u)^n]=2nr.
\ee
\Cref{eq-finite-polynomial-interpolation,eq-finite-polynomial-moment} give the numbers of interpolation points stated in \cref{sec-gaussian}. Since \(2r=\operatorname{rank}\mathsf Z_\delta\leq N\), one also has \(r\leq\lfloor N/2\rfloor\), \(2r+1\leq2\lfloor N/2\rfloor+1\), and \(2nr+1\leq2n\lfloor N/2\rfloor+1\).
\paragraph{Operator matrix elements.} For any subsystem operator \(O_A\), applying the degree-\(2r\) interpolation proved above to \(\Tr[O_AT_A(u)]\) gives
\be
\bra{\widetilde\Psi_2}(O_A\otimes\one_B)\ket{\widetilde\Psi_1}=\sum_{k=0}^{2r}L_k^{(2r)}(\ii)\,\mathcal N(u_k)\,\Tr\!\left[O_A\rho_A(u_k)\right].\label{eq-fermion-kernel-interpolation}
\ee
Setting \(O_A=\one_A\) in \cref{eq-fermion-kernel-interpolation} gives the norm kernel. Dividing the formula for general \(O_A\) by it gives the normalised off-diagonal matrix element. The Onishi formula evaluates the norm kernel, and operator kernels are conventionally evaluated by the off-diagonal Wick theorem \cite{PorroDuguet:2022Onishi}. The interpolation formula gives the same matrix elements from expectation values and normalisations of ordinary Gaussian states without either calculation.

\subsection{Bosonic Gaussian kernels, covariance, and entropy}\label{app-bosonic-gaussian-kernel-covariance}
Even finitely many bosonic modes have an infinite-dimensional Fock space, so the finite-dimensional argument in \cref{sec-entropy-continuation} does not apply. We derive \cref{eq-bosonic-transition-matrix-complex-parameters} from Gaussian kernels and give separate conditions for continuing the entropy in \cref{eq-bosonic-gaussian-entropy}.
\paragraph{Gaussian kernels and partial trace.} Let \(\mathsf X=\mathsf X^{\mathsf T}\) and \(\mathsf Y=\mathsf Y^{\mathsf T}\) be complex symmetric matrices, and set
\be
\mathsf K_\pm\coloneqq\mathsf X\pm\ii\mathsf Y.
\ee
Every square root of a determinant in this subsection is fixed by analytic continuation from its positive value at a real positive-definite matrix. Define
\be
Z(\mathsf X)\coloneqq\int_{\mathbb R^N}e^{-\bm q^{\mathsf T}\mathsf X\bm q}\,\dd^N\bm q,\qquad\langle\bm q|\rho(\mathsf X,\mathsf Y)|\bm q'\rangle\coloneqq\frac{e^{-\frac12\bm q^{\mathsf T}\mathsf K_+\bm q}e^{-\frac12{\bm q'}^{\mathsf T}\mathsf K_-\bm q'}}{Z(\mathsf X)}.\label{eq-bosonic-gaussian-kernel}
\ee
On the parameter neighbourhood under consideration, assume for every \(\bm v\in\mathbb R^N\) that
\be
\operatorname{Re}(\bm v^{\mathsf T}\mathsf X\bm v)\geq\kappa\|\bm v\|^2,\qquad\kappa>0,\qquad\operatorname{Re}(\bm v^{\mathsf T}\mathsf K_\pm\bm v)\geq\kappa_\pm\|\bm v\|^2,\qquad\kappa_\pm>0.\label{eq-bosonic-gaussian-integral-bound}
\ee
The first condition ensures convergence of the normalisation integral and all polynomial moments used below. The second ensures that \cref{eq-bosonic-gaussian-kernel} defines a trace-class rank-one operator and that the integral below equals its partial trace. Then \(Z(\mathsf X)\neq0\), and
\be
Z(\mathsf X)=\pi^{N/2}(\det\mathsf X)^{-1/2}.
\ee
For any operator \(O\) that is polynomial in the canonical operators, define
\be
\langle O\rangle_{\mathsf X,\mathsf Y}\coloneqq\frac1{Z(\mathsf X)}\int_{\mathbb R^N}e^{-\frac12\bm q^{\mathsf T}\mathsf K_-\bm q}\left[O\,e^{-\frac12\bm q^{\mathsf T}\mathsf K_+\bm q}\right]\dd^N\bm q.
\ee
This is \(\Tr[O\rho]\) for real parameters. On a neighbourhood with a uniform \(\kappa>0\), the integrands and their parameter derivatives are bounded by \(C(1+\|\bm q\|^m)e^{-\kappa\|\bm q\|^2}\) for some \(C>0\) and integer \(m\geq0\). Dominated convergence permits differentiation under the integral sign, and the normalisation and moments are holomorphic.

Write
\be
\begin{aligned} &\bm q=(\bm q_A,\bm q_B),\qquad&&\mathsf K_\pm=\begin{pmatrix}\mathsf K_{\pm,AA}&\mathsf K_{\pm,AB}\\ \mathsf K_{\pm,BA}&\mathsf K_{\pm,BB}\end{pmatrix},\\ &\mathsf C_B\coloneqq\mathsf K_{+,BB}+\mathsf K_{-,BB}=2\mathsf X_{BB},\qquad&&\bm h\coloneqq\mathsf K_{+,BA}\bm q_A+\mathsf K_{-,BA}\bm q'_A. \end{aligned}
\ee
By the first condition in \cref{eq-bosonic-gaussian-integral-bound}, \(\mathsf C_B=2\mathsf X_{BB}\) has positive real part, so the Gaussian integral over \(\bm q_B\) converges. Under the second condition, this integral is also the partial trace of the operator. Completing the square gives
\be
\langle\bm q_A|\rho_A(\mathsf X,\mathsf Y)|\bm q'_A\rangle=\frac{(2\pi)^{(N-N_A)/2}}{Z(\mathsf X)[\det\mathsf C_B]^{1/2}}\times\exp\!\left[-\frac12\bm q_A^{\mathsf T}\mathsf K_{+,AA}\bm q_A-\frac12{\bm q'_A}^{\mathsf T}\mathsf K_{-,AA}\bm q'_A+\frac12\bm h^{\mathsf T}\mathsf C_B^{-1}\bm h\right].\label{eq-bosonic-reduced-kernel}
\ee
The Gaussian integral and reduced kernel are holomorphic in the parameters while the convergence and invertibility conditions hold, so evaluating \(\mathsf X,\mathsf Y\) before or after the Gaussian integration gives the same result.

At the parameter values in \cref{eq-bosonic-complex-parameters},
\be
\mathsf K_+=\mathsf X_1+\ii\mathsf Y_1,\quad\mathsf K_-=\mathsf X_2-\ii\mathsf Y_2,\qquad\rho(\mathsf X_{\mathrm c}-\ii\mathsf Y_\delta,\mathsf Y_{\mathrm c}+\ii\mathsf X_\delta)=\frac{\ket{\Psi(\mathsf X_1,\mathsf Y_1)}\bra{\Psi(\mathsf X_2,\mathsf Y_2)}}{\braket{\Psi(\mathsf X_2,\mathsf Y_2)}{\Psi(\mathsf X_1,\mathsf Y_1)}}=\tau.
\ee
The real parts of \(\mathsf K_+\) and \(\mathsf K_-\) are \(\mathsf X_1\) and \(\mathsf X_2\), so the second condition in \cref{eq-bosonic-gaussian-integral-bound} holds. The matrix identity follows from \cref{eq-bosonic-gaussian-kernel} because the two wavefunction normalisations cancel against their overlap. Taking the partial trace and using \cref{eq-bosonic-reduced-kernel} proves \cref{eq-bosonic-transition-matrix-complex-parameters}.

For the real symmetric matrices used in \cite{Mollabashi:2021xsd}, write
\be
\mathsf W^{(a)} = \begin{pmatrix} \mathsf A^{(a)}&\mathsf B^{(a)}\\ {\mathsf B^{(a)}}^{\mathsf T}&\mathsf C^{(a)} \end{pmatrix}, \qquad\mathsf C_{\mathrm c} \coloneqq\frac{\mathsf C^{(1)}+\mathsf C^{(2)}}2, \qquad a=1,2.
\ee
Substituting \(\mathsf K_+=\mathsf W^{(1)}\), \(\mathsf K_-=\mathsf W^{(2)}\) in \cref{eq-bosonic-reduced-kernel} gives
\be
\mathsf C_B=2\mathsf C_{\mathrm c},\qquad\mathsf X_A^{(a)}=\mathsf A^{(a)}-\frac12\mathsf B^{(a)}\mathsf C_{\mathrm c}^{-1}{\mathsf B^{(a)}}^{\mathsf T},\qquad\mathsf Y_A=-\frac14\mathsf B^{(1)}\mathsf C_{\mathrm c}^{-1}{\mathsf B^{(2)}}^{\mathsf T}.
\ee
The quadratic form in the exponent of the reduced kernel is therefore
\be
-\frac12 \begin{pmatrix} \bm q_A^{\mathsf T}&{\bm q'_A}^{\mathsf T} \end{pmatrix} \begin{pmatrix} \mathsf X_A^{(1)} & 2\mathsf Y_A \\ 2\mathsf Y_A^{\mathsf T} & \mathsf X_A^{(2)} \end{pmatrix} \begin{pmatrix} \bm q_A\\ \bm q'_A \end{pmatrix}.
\ee
The matrices \(\mathsf X_A^{(1)},\mathsf X_A^{(2)}\), and \(\mathsf Y_A\) agree with those in the reduced kernel in \cite{Mollabashi:2021xsd}. For these \(\mathsf K_\pm\), \cref{eq-bosonic-reduced-kernel} gives the Gaussian kernel of the transition matrix used there and the reduced kernel in \cref{eq-bosonic-transition-matrix-complex-parameters}.
\paragraph{Covariance.} Define \(\bm p\coloneqq(p_1,\ldots,p_N)^{\mathsf T}\), with \(p_j=-\ii\partial_{q_j}\). Under the first condition in \cref{eq-bosonic-gaussian-integral-bound}, integration by parts and \([q_i,p_j]=\ii\delta_{ij}\) give
\begin{align}
&0=\frac1{Z(\mathsf X)}\int_{\mathbb R^N}\partial_{q_i}\!\left[q_j e^{-\bm q^{\mathsf T}\mathsf X\bm q}\right]\dd^N\bm q =\delta_{ij}-2\sum_k\mathsf X_{ik}\langle q_jq_k\rangle_{\mathsf X,\mathsf Y},\\ &p_j e^{-\bm q^{\mathsf T}\mathsf K_+\bm q/2} =\ii(\mathsf K_+\bm q)_j e^{-\bm q^{\mathsf T}\mathsf K_+\bm q/2}, \quad p_ip_j e^{-\bm q^{\mathsf T}\mathsf K_+\bm q/2} =\left[(\mathsf K_+)_{ij}-(\mathsf K_+\bm q)_i(\mathsf K_+\bm q)_j\right]e^{-\bm q^{\mathsf T}\mathsf K_+\bm q/2},\\ &\left\langle\bm q\bm q^{\mathsf T}\right\rangle_{\mathsf X,\mathsf Y} =\frac12\mathsf X^{-1},\quad\frac12\left\langle q_ip_j+p_jq_i\right\rangle_{\mathsf X,\mathsf Y} =-\frac12(\mathsf X^{-1}\mathsf Y)_{ij}, \quad\left\langle\bm p\bm p^{\mathsf T}\right\rangle_{\mathsf X,\mathsf Y} =\frac12\left(\mathsf X+\mathsf Y\mathsf X^{-1}\mathsf Y\right).
\end{align}
The first moments vanish by oddness. Hence the entries of \(\Gamma\) are holomorphic, and \cref{eq-bosonic-covariance-matrix} continues to hold for complex parameters satisfying the first condition in \cref{eq-bosonic-gaussian-integral-bound}. Partial trace preserves moments of operators supported in \(A\), so \(\Gamma_A\) is obtained by retaining the rows and columns belonging to \(q_1,\ldots,q_{N_A},p_1,\ldots,p_{N_A}\).
\paragraph{Entropy.} Use the eigenvalue branches \(\nu_j\) selected before \cref{eq-bosonic-gaussian-entropy}, omitting any branch that is identically \(1/2\), since its entropy contribution is identically zero. Assume that the group of eigenvalues continued from the remaining positive eigenvalues stays separated from the other eigenvalues, and choose a fixed positively oriented contour \(\mathcal C_+\) enclosing only this group, with each eigenvalue enclosed exactly once. The Riesz projection is holomorphic while this separation persists \cite{Kato:1995perturbation}.

If \(\mathcal C_+\) and its interior lie in \(\mathbb C\setminus(-\infty,1/2]\), then \cref{eq-matrix-function-contour} gives
\begin{align}
S(\Gamma_A)=\frac{1}{2\pi\ii}\oint_{\mathcal C_+} &\left[ \left(\zeta+\frac12\right)\Log\!\left(\zeta+\frac12\right) -\left(\zeta-\frac12\right)\Log\!\left(\zeta-\frac12\right) \right]\\[-2mm] &\times\Tr\!\left[\left(\zeta\one-\ii J_A\Gamma_A\right)^{-1}\right]\dd\zeta.
\end{align}
For real parameters this is \cref{eq-bosonic-gaussian-entropy}. The contour formula is holomorphic in the parameters and allows degeneracies in the selected group of eigenvalues. For a Taylor expansion, we assume the same conditions on spectral separation and contours throughout its disc of convergence.

\subsection{Polynomial degree of the bosonic covariance}\label{app-bosonic-covariance-polynomial-degree}
The polynomial degree of \(\det\mathsf X(u)\,\Gamma_A(u)\) gives a sufficient number of ordinary covariance matrices for \cref{eq-bosonic-covariance-interpolation}. We prove the degree bounds and the nonvanishing of \(\det\mathsf X(\ii)\) needed for division in this formula.
\paragraph{Degree bound.} For the family in \cref{eq-bosonic-interpolation-family}, write \(\operatorname{adj}\) for the adjugate matrix. Then \cref{eq-bosonic-covariance-matrix} gives
\be
\begin{aligned} &2\det\mathsf X(u)\,\Gamma(u) = \left(\begin{array}{@{}l@{\qquad}l@{}} \operatorname{adj}\mathsf X(u) & -\operatorname{adj}\mathsf X(u)\mathsf Y(u) \\ -\mathsf Y(u)\operatorname{adj}\mathsf X(u) & \det\mathsf X(u)\mathsf X(u) + \mathsf Y(u)\operatorname{adj}\mathsf X(u)\mathsf Y(u) \end{array}\right), \\ &\mathsf L \coloneqq\mathsf X_{\mathrm c}^{-1/2} \mathsf Y_\delta\mathsf X_{\mathrm c}^{-1/2}, \qquad\mathsf X(u) = \mathsf X_{\mathrm c}^{1/2} (\one-u\mathsf L) \mathsf X_{\mathrm c}^{1/2}, \qquad\operatorname{rank}\mathsf L = r_{\mathrm b}. \end{aligned}
\ee
Since \(\mathsf L\) is real symmetric, write \(\mathsf L=\mathsf O\operatorname{diag}(\lambda_1,\ldots,\lambda_{r_{\mathrm b}},0,\ldots,0)\mathsf O^{\mathsf T}\) with \(\mathsf O\) orthogonal. Then \(\det(\one-u\mathsf L)=\prod_{a=1}^{r_{\mathrm b}}(1-u\lambda_a)\), while each entry of \(\operatorname{adj}(\one-u\mathsf L)\) is a linear combination of products obtained by omitting one diagonal factor. Multiplication by the constant matrices \(\mathsf X_{\mathrm c}^{1/2}\) does not change these degree bounds. The four blocks satisfy the following degree bounds in \(u\)
\be
\begin{array}{@{}l@{\qquad}l@{}} \deg\det\mathsf X = r_{\mathrm b}, & \deg\operatorname{adj}\mathsf X = \min(N-1,r_{\mathrm b}), \\ \multicolumn{2}{@{}l@{}}{\deg[\operatorname{adj}\mathsf X\,\mathsf Y], \ \deg[\mathsf Y\,\operatorname{adj}\mathsf X] \leq\min(N-1,r_{\mathrm b})+1,} \\ \deg[ \det\mathsf X\,\mathsf X + \mathsf Y\,\operatorname{adj}\mathsf X\,\mathsf Y ] \leq d_{\mathrm b}^{\max}, & \deg[\det\mathsf X\,\Gamma_A] \leq d_{\mathrm b}^{\max} = \min(N+1,r_{\mathrm b}+2). \end{array}
\ee
Write \(\widetilde\Gamma_A(u)=\sum_{k=0}^{d_{\mathrm b}^{\max}}\widetilde\Gamma_{A,k}u^k\). Its degree is \(d_{\mathrm b,A}\) as defined in \cref{eq-bosonic-covariance-degree}. The number \(d_{\mathrm b}^{\max}\) is an upper bound before checking which coefficients vanish after restriction to \(A\).
\be
\operatorname{spec}(\one-\ii\mathsf L)=\{1-\ii\lambda\mid\lambda\in\operatorname{spec}\mathsf L\}\subset\mathbb C\setminus\{0\}\quad\Longrightarrow\quad\det\mathsf X(\ii)\neq0.\label{eq-bosonic-interpolation-determinant}
\ee
By \cref{eq-lagrange-polynomials,eq-bosonic-interpolation-determinant}, interpolating this degree-\(d_{\mathrm b,A}\) polynomial at \(d_{\mathrm b,A}+1\) distinct points proves \cref{eq-bosonic-covariance-interpolation}.
\paragraph{The case \(\mathsf Y_1=\mathsf Y_2\).}
\be
\mathsf Y_1=\mathsf Y_2\Rightarrow\mathsf Y_\delta=0,\quad\mathsf X(u)=\mathsf X_{\mathrm c},\qquad\ket{\Psi(u)}\coloneqq\ket{\Psi(\mathsf X(u),\mathsf Y(u))}=\exp\!\left[-\frac{\ii u}{2}\bm q^{\mathsf T}\mathsf X_\delta\bm q\right]\ket{\Psi(0)}.
\ee
Let \(\mathsf E_A\in\mathbb R^{N_A\times N}\) select the position coordinates belonging to subsystem \(A\). The quadratic coefficient of the restricted momentum block is
\be
[u^2](2\Gamma_A(u))_{pp}=\mathsf E_A\mathsf X_\delta\mathsf X_{\mathrm c}^{-1}\mathsf X_\delta\mathsf E_A^{\mathsf T}.
\ee
It is positive semidefinite and vanishes only if \(\mathsf E_A\mathsf X_\delta=0\). In that case the linear coefficients of the restricted \(qp\) and \(pq\) blocks and the linear and quadratic coefficients of the restricted \(pp\) block also vanish, so the covariance is independent of \(u\). Otherwise \(d_{\mathrm b,A}=2\), as stated after \cref{eq-bosonic-three-covariance-interpolation,eq-bosonic-three-covariance-real-imaginary}.

\subsection{One harmonic oscillator}\label{app-oscillator}
The oscillator identities in \cref{eq-oscillator-complex-time-identities} give the matrix and covariance formulae for several independent modes in \cref{eq-multimode-complex-time-formulae}. We prove these identities and their frequency conditions, which are not required for \cref{eq-bosonic-transition-matrix-complex-parameters}.
The normalised ground-state wavefunction and overlap are
\be
\langle q|0_\omega\rangle=\left(\frac{\omega}{\pi}\right)^{1/4}e^{-\omega q^2/2},\qquad\braket{0_{\omega_1}}{0_{\omega_2}}=\left(\frac{2\sqrt{\omega_1\omega_2}}{\omega_1+\omega_2}\right)^{1/2}>0.\label{eq-oscillator-vacuum}
\ee
If \(\omega=\omega_1=\omega_2\), define
\be
(\lambda_\Omega,\theta_\omega,\Omega)=(0,0,\omega).
\ee
The formula for \(\theta_\omega\) in \cref{eq-oscillator-frequency-and-time-shift} is undefined when all three frequencies are equal and is used only when \(\lambda_1\lambda_2>0\).
\paragraph{Operator identity.} Let \(a_\omega\) be the annihilation operator for \(H_\omega\), so that \(H_\omega=\omega(a_\omega^\dagger a_\omega+\tfrac12)\). Using \cref{eq-oscillator-vacuum,eq-oscillator-frequency-and-time-shift}, conjugation by \(e^{-\theta H_\omega}\) multiplies the coefficient of \(a_\omega^{\dagger2}\) by \(e^{-2\omega\theta}\). Therefore
\be
\ket{0_{\omega_i}}\propto\exp\!\left(\frac{\lambda_i}{2}a_\omega^{\dagger2}\right)\ket{0_\omega},\quad e^{-\theta H_\omega}a_\omega^\dagger e^{+\theta H_\omega}=e^{-\omega\theta}a_\omega^\dagger,\quad\lambda_1=\lambda_\Omega e^{-2\omega\theta_\omega},\thickspace\lambda_2=\lambda_\Omega e^{+2\omega\theta_\omega}.\label{eq-oscillator-squeezed-state-coefficients}
\ee
To prove necessity, write any ordinary zero-mean pure Gaussian state in the expansion in \cref{eq-oscillator-squeezed-state-coefficients}, with coefficient \(\lambda_0\), where \(|\lambda_0|<1\). A finite imaginary-time shift \(\theta\in\mathbb R\) generated by \(H_\omega\) gives the required ket and bra coefficients only if
\be
\lambda_1=\lambda_0e^{-2\omega\theta},\qquad\lambda_2=\overline{\lambda_0}e^{2\omega\theta},\qquad\lambda_1\lambda_2=|\lambda_0|^2.
\ee
Thus \(\lambda_1\lambda_2<0\) is impossible. If exactly one \(\lambda_i\) vanishes, no finite \(\lambda_0\) and \(\theta\) satisfy these equations. If both vanish, then \(\omega=\omega_1=\omega_2\). Conversely, \(\lambda_1\lambda_2>0\) gives the real \(\lambda_\Omega\), \(\theta_\omega\), and \(\Omega\) in \cref{eq-oscillator-frequency-and-time-shift}. Apart from the case of three equal frequencies, \cref{eq-oscillator-frequency-condition} is necessary and sufficient for writing the transition matrix as \(\rho^{(\Omega)}(t-\ii\theta_\omega)\) with \(\Omega>0\) and finite real \(\theta_\omega\). It is not a restriction on the analytic continuation in \cref{eq-bosonic-transition-matrix-complex-parameters}. By \cref{eq-oscillator-squeezed-state-coefficients,eq-oscillator-vacuum}, for nonzero real \(C_1,C_2\),
\be
\begin{aligned} &e^{-\theta_\omega H_\omega}\ket{0_\Omega} = C_1\ket{0_{\omega_1}}, & \qquad&e^{+\theta_\omega H_\omega}\ket{0_\Omega} = C_2\ket{0_{\omega_2}}, \\ &1 = C_1C_2\braket{0_{\omega_2}}{0_{\omega_1}}, & &e^{-\theta_\omega H_\omega} \proj{0_\Omega} e^{+\theta_\omega H_\omega} = \frac{ \ket{0_{\omega_1}}\bra{0_{\omega_2}} }{ \braket{0_{\omega_2}}{0_{\omega_1}} }. \end{aligned}\label{eq-oscillator-ground-state-identity}
\ee
Conjugating \cref{eq-oscillator-ground-state-identity} by \(e^{-\ii tH_\omega}\) and \(e^{+\ii tH_\omega}\) proves the second identity in \cref{eq-oscillator-complex-time-identities}. The third follows by taking second moments.

Although \(e^{+\theta_\omega H_\omega}\) is unbounded for one sign of \(\theta_\omega\), both vectors in \cref{eq-oscillator-ground-state-identity} belong to the Hilbert space because
\be
|\lambda_\Omega|e^{2\omega|\theta_\omega|}=\max(|\lambda_1|,|\lambda_2|)<1.
\ee
The rank-one product is trace class. When \(\lambda_1\lambda_2>0\), the Fock-space expansion of the squeezed state is holomorphic on
\be
|\operatorname{Im}s|<\frac1{2\omega}\log\frac1{|\lambda_\Omega|}.
\ee
The value \(s=t-\ii\theta_\omega\) lies in this strip. If \(\omega=\omega_1=\omega_2\), \(\rho^{(\Omega)}(s)\) is independent of \(s\) and hence entire.
\paragraph{Covariance entries.} Define
\begin{align}
&X^{1|2}(t)\coloneqq\Tr[q^2\tau(t)],\quad P^{1|2}(t)\coloneqq\Tr[p^2\tau(t)],\quad R^{1|2}(t)\coloneqq\frac12\Tr[(qp+pq)\tau(t)], \\ &\omega_{\mathrm c}\coloneqq\frac{\omega_1+\omega_2}{2},\quad\omega_\delta\coloneqq\frac{\omega_2-\omega_1}{2},\quad X^{1|2}(0)=\frac1{2\omega_{\mathrm c}},\quad P^{1|2}(0)=\frac{\omega_1\omega_2}{2\omega_{\mathrm c}},\quad R^{1|2}(0)=-\frac{\ii\omega_\delta}{2\omega_{\mathrm c}}, \\ &q(t)=q\cos(\omega t)+\frac p\omega\sin(\omega t),\qquad p(t)=p\cos(\omega t)-\omega q\sin(\omega t).
\end{align}
The values at \(t=0\) follow from \cref{eq-oscillator-vacuum,eq-oscillator-transition-matrix}, and the formulae for \(q(t)\) and \(p(t)\) give the Heisenberg evolution. Therefore
\begin{align}
&X^{1|2}(t) = \frac{ \omega^2\cos^2(\omega t) +\omega_1\omega_2\sin^2(\omega t) -\ii\omega\omega_\delta\sin(2\omega t) }{ 2\omega^2\omega_{\mathrm c} },\label{eq-oscillator-position-covariance} \\ &P^{1|2}(t) = \frac{ \omega^2\sin^2(\omega t) +\omega_1\omega_2\cos^2(\omega t) +\ii\omega\omega_\delta\sin(2\omega t) }{ 2\omega_{\mathrm c} }, \\ &R^{1|2}(t) = \frac{ (\omega_1\omega_2-\omega^2)\sin(2\omega t) -2\ii\omega\omega_\delta\cos(2\omega t) }{ 4\omega\omega_{\mathrm c} }.\label{eq-oscillator-mixed-covariance}
\end{align}
For \(\rho^{(\Omega)}(t)=e^{-\ii tH_\omega}\proj{0_\Omega}e^{+\ii tH_\omega}\), define
\be
\bigl(X^{(\Omega)},P^{(\Omega)},R^{(\Omega)}\bigr)(t)\coloneqq\left.\bigl(X^{1|2},P^{1|2},R^{1|2}\bigr)(t)\right|_{\omega_1=\omega_2=\Omega},\qquad\omega_{\mathrm c}=\Omega,\quad\omega_\delta=0.
\ee
Setting \(\omega_1=\omega_2=\Omega\) and then replacing \(t\) by \(t-\ii\theta_\omega\) gives the right-hand sides of \crefrange{eq-oscillator-position-covariance}{eq-oscillator-mixed-covariance} for the original \(\omega_1,\omega_2\).

\subsection{Real normal modes of the scalar lattice}\label{app-scalar-real-modes}
Real normal modes give independent harmonic oscillators, so the oscillator identities give \cref{eq-scalar-transition-matrix-complex-times} at finite \(N\) under \cref{eq-scalar-frequency-condition}. The covariance in site coordinates then gives the subsystem entropy for the scalar comparison in \cref{fig-quench-comparison}.
For the periodic \(N\)-site lattice in \cref{sec-scalar} \cite{MohammadiMozaffar:2017nri,He:2017wla,Mollabashi:2021xsd}, with \(z\in\mathbb Z_{>0}\) and site labels understood modulo \(N\),
\be
H(m,z)=\frac12\sum_{a=0}^{N-1}\left[p_a^2+m^{2z}q_a^2+\left(\sum_{k=0}^{z}(-1)^{z+k}\binom zkq_{a-1+k}\right)^2\right].\label{eq-lattice-position-Hamiltonian}
\ee
Use
\begin{align}
&Q_n\coloneqq\frac1{\sqrt N} \sum_{a=0}^{N-1} e^{-2\pi\ii na/N}q_a, & \qquad&\Pi_n\coloneqq\frac1{\sqrt N} \sum_{a=0}^{N-1} e^{-2\pi\ii na/N}p_a,\label{eq-lattice-DFT} \\ &Q_n^\dagger= Q_{-n}, & &\Pi_n^\dagger= \Pi_{-n}, \qquad[Q_n,\Pi_{n'}] = \ii\delta_{n+n',\,0\thickspace(\mathrm{mod}\,N)}.
\end{align}
The unitary discrete Fourier transform in \cref{eq-lattice-DFT} has the same plane-wave phase convention as \cite{Mollabashi:2021xsd}. Substituting \cref{eq-lattice-DFT} into \cref{eq-lattice-position-Hamiltonian} gives
\be
H(m,z)=\frac12\sum_{n=0}^{N-1}\left[\Pi_n\Pi_{-n}+\omega_n(m,z)^2Q_nQ_{-n}\right].\label{eq-lattice-Fourier-modes}
\ee
For one representative of each pair \(n\neq-n\pmod N\), define
\be
q_{n,\mathrm c}\coloneqq\frac{Q_n+Q_{-n}}{\sqrt2},\qquad p_{n,\mathrm c}\coloneqq\frac{\Pi_n+\Pi_{-n}}{\sqrt2},\qquad q_{n,\mathrm s}\coloneqq\frac{Q_n-Q_{-n}}{\ii\sqrt2},\qquad p_{n,\mathrm s}\coloneqq\frac{\Pi_n-\Pi_{-n}}{\ii\sqrt2}.\label{eq-lattice-real-coordinates}
\ee
These are two independent real canonical pairs of frequency \(\omega_n(m,z)\). There is one pair for \(n=0\) and one for \(n=N/2\) when \(N\) is even. Substituting \cref{eq-lattice-real-coordinates} into \cref{eq-lattice-Fourier-modes} gives the Hamiltonian in \cref{eq-lattice-real-mode-Hamiltonian}, so the ground state factorises over the real normal modes.

Under \cref{eq-scalar-frequency-condition}, every real mode satisfies \cref{eq-oscillator-frequency-condition}. Applying \cref{eq-oscillator-complex-time-identities} to every real normal mode and taking the tensor product proves \cref{eq-scalar-transition-matrix-complex-times} at finite \(N\). Passing to the infinite-lattice or continuum limit requires a regulated limit that converges locally uniformly on the complex domain and separate control of any vanishing total overlap.
\paragraph{Covariance in site coordinates.} Different real modes have vanishing cross-correlations, and the cosine and sine modes associated with the same \(n\) have equal covariances. Thus
\be
\begin{aligned} &\langle Q_nQ_{n'}\rangle= \delta_{n+n',\,0\thickspace(\mathrm{mod}\,N)}X_n, & \qquad&\langle\Pi_n\Pi_{n'}\rangle= \delta_{n+n',\,0\thickspace(\mathrm{mod}\,N)}P_n, \\ &\frac12 \langle Q_n\Pi_{n'}+\Pi_{n'}Q_n \rangle= \delta_{n+n',\,0\thickspace(\mathrm{mod}\,N)}R_n. \end{aligned}\label{eq-lattice-mode-covariance}
\ee
For each \(n\), let \(\Omega_n,\theta_n\) be obtained from \(\omega_{1,n},\omega_{2,n},\omega_n\) through \cref{eq-oscillator-frequency-and-time-shift}. Taking second moments of \cref{eq-scalar-transition-matrix-complex-times} gives
\be
X_n^{1|2}(t)=X_n^{(\Omega)}(t-\ii\theta_n),\qquad P_n^{1|2}(t)=P_n^{(\Omega)}(t-\ii\theta_n),\qquad R_n^{1|2}(t)=R_n^{(\Omega)}(t-\ii\theta_n),\label{eq-scalar-mode-complex-times}
\ee
where the right-hand sides are obtained from \crefrange{eq-oscillator-position-covariance}{eq-oscillator-mixed-covariance} by setting \(\omega_1=\omega_2=\Omega_n\) and \(\omega=\omega_n\). Applying the inverse of \cref{eq-lattice-DFT} to \cref{eq-lattice-mode-covariance} gives
\be
F_{ab}^{1|2}(t)=\frac1N\sum_{n=0}^{N-1}e^{2\pi\ii n(a-b)/N}F_n^{1|2}(t),\qquad F\in\{X,P,R\}.
\ee
Together with \cref{eq-scalar-mode-complex-times,eq-lattice-mode-covariance}, these inverse Fourier transforms give the covariance in site coordinates. Restricting it to \(A\) and applying \cref{eq-bosonic-gaussian-entropy} gives the comparison stated in \cref{sec-scalar}.
\bibliographystyle{JHEP}
\bibliography{refs}
\end{document}